%% file: main.tex
\documentclass{ieeeaccess}
\usepackage[T1]{fontenc}
\usepackage[utf8]{inputenc}
\usepackage{cite}
\usepackage{amsmath,amssymb,amsfonts,graphicx,booktabs,array,textcomp}
\usepackage{etoolbox,placeins}
\usepackage[hidelinks,hypertexnames=false]{hyperref}
\newcommand{\tightlist}{}
\def\headerlogo{\raisebox{-2pt}{\includegraphics[width=7.61pc]{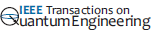}}}
\def\headerlogoall{\raisebox{-2pt}{\includegraphics[width=7.61pc]{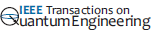}}}
\makeatletter
\pretocmd{\@makecaption}{\def\xfigwd{0pt}}{}{}
\patchcmd{\@maketitle}{Digital Object Identifier\space\@doi}{}{}{\PackageError{tqe-draft}{DOI patch failed}{Check the template version.}}
\apptocmd{\ps@titlepage}{\def\@oddfoot{\hfil\footerpagefont\thepage\hfil}\let\@evenfoot\@oddfoot}{}{}
\apptocmd{\ps@headings}{\def\@oddfoot{\hfil\footerpagefont\thepage\hfil}\let\@evenfoot\@oddfoot}{}{}
\makeatother
\begin{document}
\history{}
\doi{}
\title{Scaling Laws for Parameter Setting in Digitized Counterdiabatic QAOA}
\author{\uppercase{Yaocheng Chen}\authorrefmark{1} and \uppercase{Jiapei Zhuang}\authorrefmark{2}}
\address[1]{Instituto de Astronomia, Geof\'isica e Ci\^encias Atmosf\'ericas (IAG-USP), Universidade de S\~ao Paulo, S\~ao Paulo 05508-090, Brazil}
\address[2]{Quantum Science Center of Guangdong-Hong Kong-Macao Greater Bay Area, Shenzhen 518045, China}
\corresp{Corresponding authors: Yaocheng Chen (e-mail: \href{mailto:yaochengchen@usp.br}{yaochengchen@usp.br}) and Jiapei Zhuang (e-mail: \href{mailto:jpzhuangphy@gmail.com}{jpzhuangphy@gmail.com}).}
\tfootnote{This study was financed, in part, by the S\~ao Paulo Research Foundation (FAPESP), Brasil. Process Numbers \#2025/25586-5 and \#2021/02120-0.}
\markboth{Chen and Zhuang: Scaling Laws for Parameter Setting in Digitized Counterdiabatic QAOA}{Chen and Zhuang: Scaling Laws for Parameter Setting in Digitized Counterdiabatic QAOA}
\input{abstract.tex}
\titlepgskip=-15pt
\maketitle
\pagestyle{headings}
\normalfont
\raggedbottom

\input{body.tex}

\eoddefinedtrue
\end{document}


\begin{center}
{\Large Supplementary Material}\\[6pt]
Scaling Laws for Parameter Setting in Digitized Counterdiabatic QAOA\\[4pt]
Yaocheng Chen and Jiapei Zhuang
\end{center}

\section*{S1. First-layer angle rescaling at depth four}
\begin{figure}[htbp]
\centering
\includegraphics[width=0.95\linewidth]{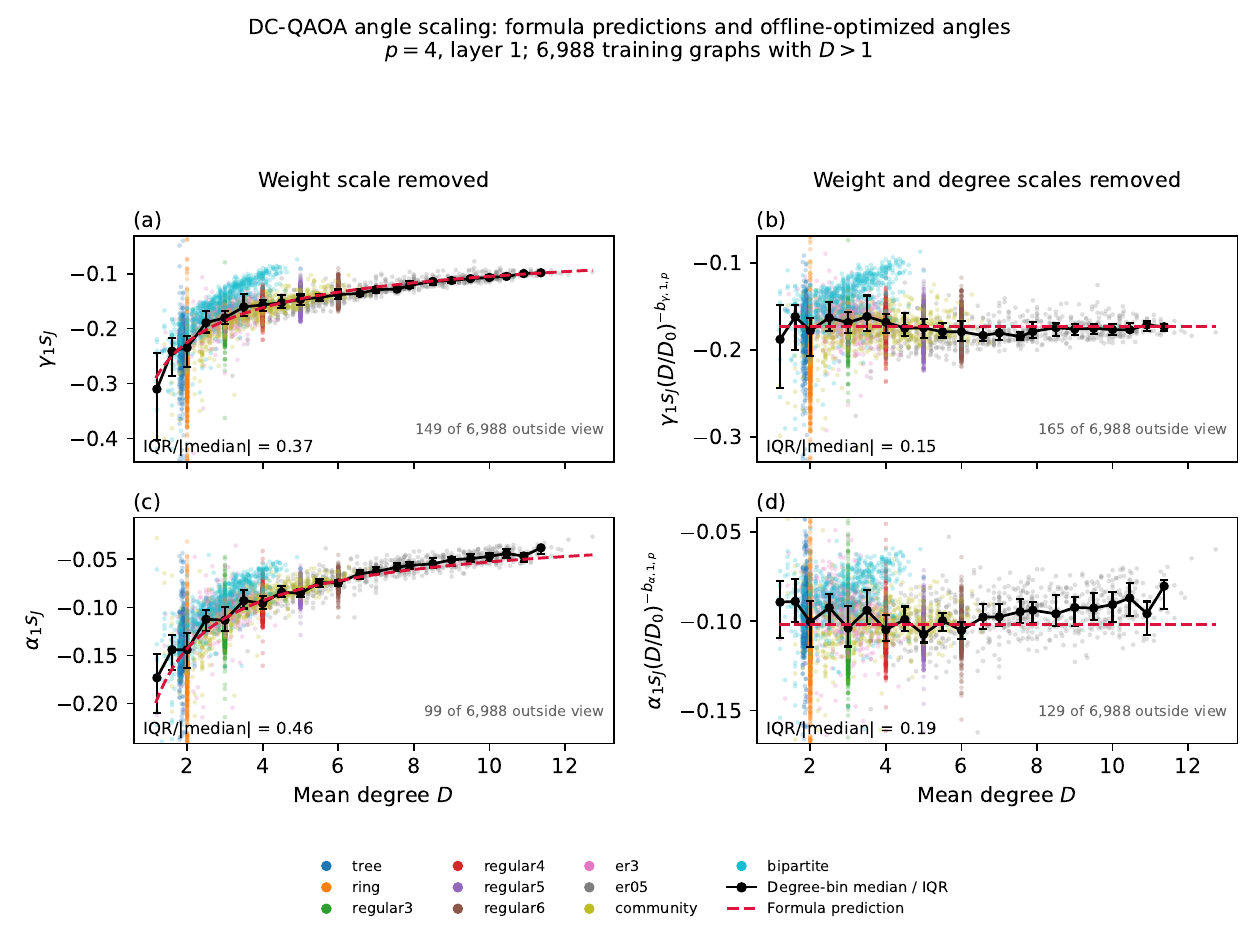}
\caption{
DC-QAOA angle scaling: formula predictions and offline-optimized angles for the first layer at $p=4$, using 6988 training graphs with $D>1$. The upper and lower rows show cost and counterdiabatic angles. The left column shows $\gamma_1s_J$ and $\alpha_1s_J$; the right column shows $\gamma_1s_J(D/D_0)^{-b_{\gamma,1,4}}$ and $\alpha_1s_J(D/D_0)^{-b_{\alpha,1,4}}$. Points are colored by graph family. Black markers and error bars show degree-bin medians and interquartile ranges (bin width 0.5; at least 10 records). Red dashed lines show the depth-4 first-layer power laws on the left and their fitted amplitudes on the right. All 6988 records contribute to the dispersion statistics, including points outside the displayed ranges. The 12 low-degree records are excluded from this fitting-range diagnostic.}\label{fig:depth4-rescaling}
\end{figure}

Degree rescaling reduces IQR/$|\mathrm{median}|$ from 0.37 to 0.15 for the first-layer cost angle and from 0.46 to 0.19 for the counterdiabatic angle. The rescaled pooled medians are $-0.1731$ and $-0.1021$, compared with fitted amplitudes of $-0.1735$ and $-0.1019$. The exponents are $b_{\gamma,1,4}=-0.4738$ and $b_{\alpha,1,4}=-0.6182$. The same rescaling therefore narrows the first-layer distributions at both $p=1$ and $p=4$, with residual variation across graph instances.
\clearpage
\section*{S2. Effective degree exponents of first-layer angles}
\begingroup\small
We estimate effective degree exponents separately at each circuit depth from the first-layer training angles, using the same sign convention as in the main text. After restricting the records to \(D>1\), we group \(y=\alpha_1s_J\) or \(y=\gamma_1s_J\) into degree bins of width 0.5, retaining bins with at least 10 records. Bin membership is assigned by rounding \(D/0.5\) to the nearest integer, with ties rounded to the nearest even integer, and multiplying by 0.5. For each bin, we take the median degree and the median angle. 
We fit $\log|m_k|$ against $\log(d_k/D_0)$, where $m_k$ and $d_k$ are the angle and degree medians in bin $k$, weighting squared residuals by the bin count. Bins with zero angle median are excluded. The fixed reference $D_0$ shifts the intercept but leaves the fitted slope and its bootstrap interval unchanged. The slope defines the effective exponent. The 95\% percentile intervals use 500 graph-bootstrap resamples, with binning and fitting repeated in each resample. The random seed is 20261024.

For \(p=1,2,3,4\), filtering the 7000 training records to \(D>1\) leaves 6988 records per depth. At \(p=6\), the archived pool contains 4000 records from 10--16-qubit graphs and 1000 from 18-qubit graphs; the same filter leaves 4988. Thus, the \(p=6\) point uses a different size composition. These are the records available before the bin-count threshold is applied.

\begin{figure}[htbp]
\centering
\includegraphics[width=0.72\linewidth]{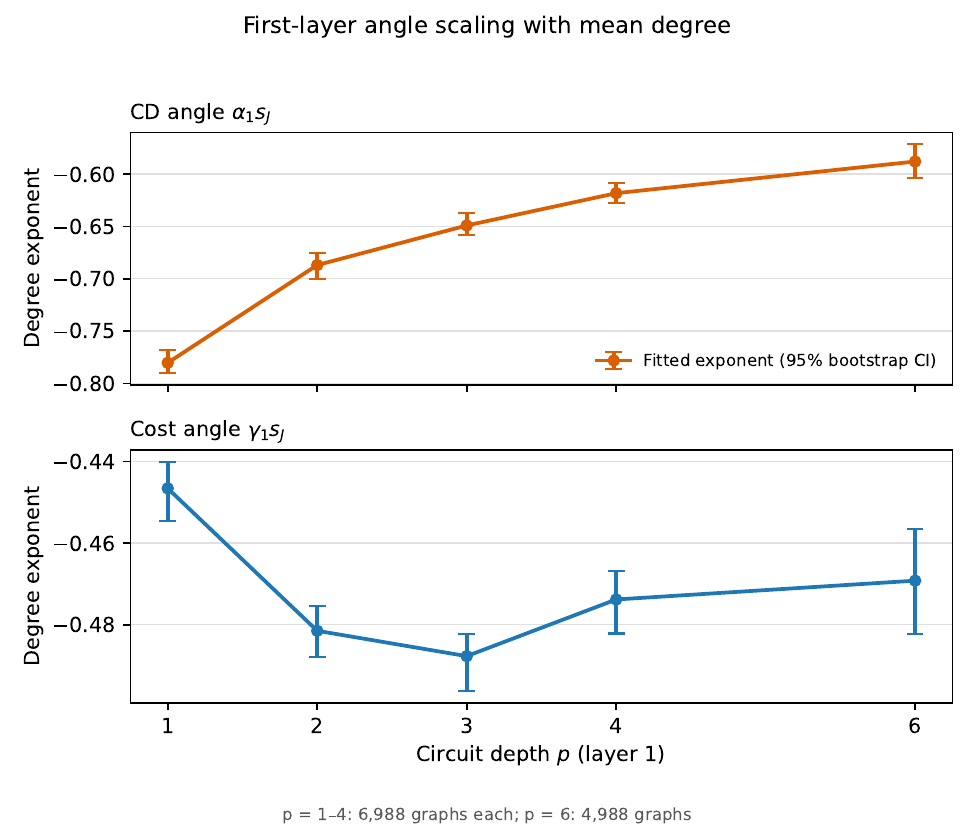}
\caption{
First-layer degree exponents of the power-law parameter-setting rule. The upper and lower panels show the exponents fitted to $\alpha_1s_J$ and $\gamma_1s_J$, respectively. Markers show fitted slopes; error bars show 95\% graph-bootstrap percentile intervals from 500 resamples. Lines connect discrete depths to guide the eye. Record counts refer to the $D>1$ selection before bin filtering.}\label{fig:effective-exponents}
\end{figure}

The fitted first-layer counterdiabatic exponent changes from approximately $-0.780$ at $p=1$ to $-0.588$ at $p=6$. Cost-angle exponents range from approximately $-0.488$ to $-0.447$. 
The power-law rule uses these depth-specific first-layer fits; later-layer coefficients are fitted separately, as listed in Table~1 of the main text. The intervals quantify uncertainty in the training fits, not in transferred performance.
\endgroup
\clearpage
\begingroup

\section*{S3. Hardware execution and paired results}
\label{supp:hardware}
\subsection*{Instances and frozen parameters}
The hardware evaluation uses the 21 unit-weight 3-regular graphs in the balanced 1050-graph sample, comprising three graphs at each $n=10,12,14,16,18,20,22$. The graph IDs and angle vectors are recorded in \texttt{tqe\_hw\_plan.json}. Both methods use the same graph instances at $p=1$ and $2$, giving 84 circuits. The power-law coefficients and the cohort- and depth-specific QAOA schedules are those used in the ideal benchmarks; neither method is optimized on the hardware test instances.

\subsection*{Circuit compilation and execution}
The circuits follow the gate ordering and angle conventions in main-text Section~\mbox  {II-B}, followed by computational-basis measurement of all logical qubits. They were transpiled for \texttt{ibm\_fez} at optimization level 3 with \texttt{seed\_transpiler=1}. The recorded software versions are Qiskit 2.6.1 \cite{qiskit2024} and Qiskit IBM Runtime 0.51.0. Each circuit was executed with 8000 shots in a single job on 28 October 2026. Dynamical decoupling, gate twirling, and measurement twirling were enabled; the result metadata record 32 randomizations per circuit. The reported probabilities use the returned logical-bit counts without additional offline readout-error mitigation or zero-noise extrapolation.

Transpilation can assign different physical layouts and routing operations to the two methods, so the comparison is between their compiled implementations.

\subsection*{Probability reconstruction and aggregation}
For each graph, classical enumeration identifies all bit strings attaining the maximum cut value. The hardware estimate is the total observed count of these bit strings divided by 8000, including all degenerate optima. Classical bit $i$ is interpreted as bit $i$ of the little-endian integer used to evaluate the graph energy. Measurement instructions and final layouts were checked for all 84 circuits. For the 24 circuits at $n=10,12$, the ideal logical output distributions of the transpiled circuits were also checked against their untranspiled counterparts. The ideal reference probabilities for all sizes are taken from the corresponding PennyLane benchmark records.

Size-resolved means average the three graphs equally. Overall means average all 21 graphs equally, and paired differences compare the two methods on the same graph at the same depth. Fig.~\ref{fig:hardware-paired} shows every graph. No confidence intervals are assigned to the hardware results, because finite shots, variation between randomizations, and device drift are not captured by the graph bootstrap used for the ideal benchmarks.

\begin{table}[H]
\centering

\caption{Raw hardware summary, averaging 21 graphs at each depth. Probabilities and paired differences are in percent and percentage points, respectively. PL denotes power-law DC-QAOA; Q denotes fixed-angle QAOA. Higher/tied/lower counts refer to PL relative to Q}
\label{tab:hardware-summary}
\begin{tabular}{c c c c c}
\hline
$p$ & $P_{\rm opt}^{\rm PL}$ (\%) & $P_{\rm opt}^{\rm Q}$ (\%) & Difference (pp) & Higher/tied/lower\\
\hline
1 & 1.61 & 0.76 & $+0.85$ & 20/0/1\\
2 & 1.25 & 1.61 & $-0.36$ & 4/1/16\\
\hline
\end{tabular}
\end{table}

\subsection*{Resources and interpretation}
Mean transpiled two-qubit gate counts are 228.8 and 108.8 for PL and Q at $p=1$, and 479.7 and 232.6 at $p=2$. Fig.~\ref{fig:hardware-resources}(b) reports the ratio of the three-graph mean hardware probability to the corresponding mean ideal probability. This is a ratio of means and not a fidelity estimate. The results come from one device and one execution job, with three graphs per size.

\begin{figure}[H]
\centering

\includegraphics[width=0.98\linewidth]{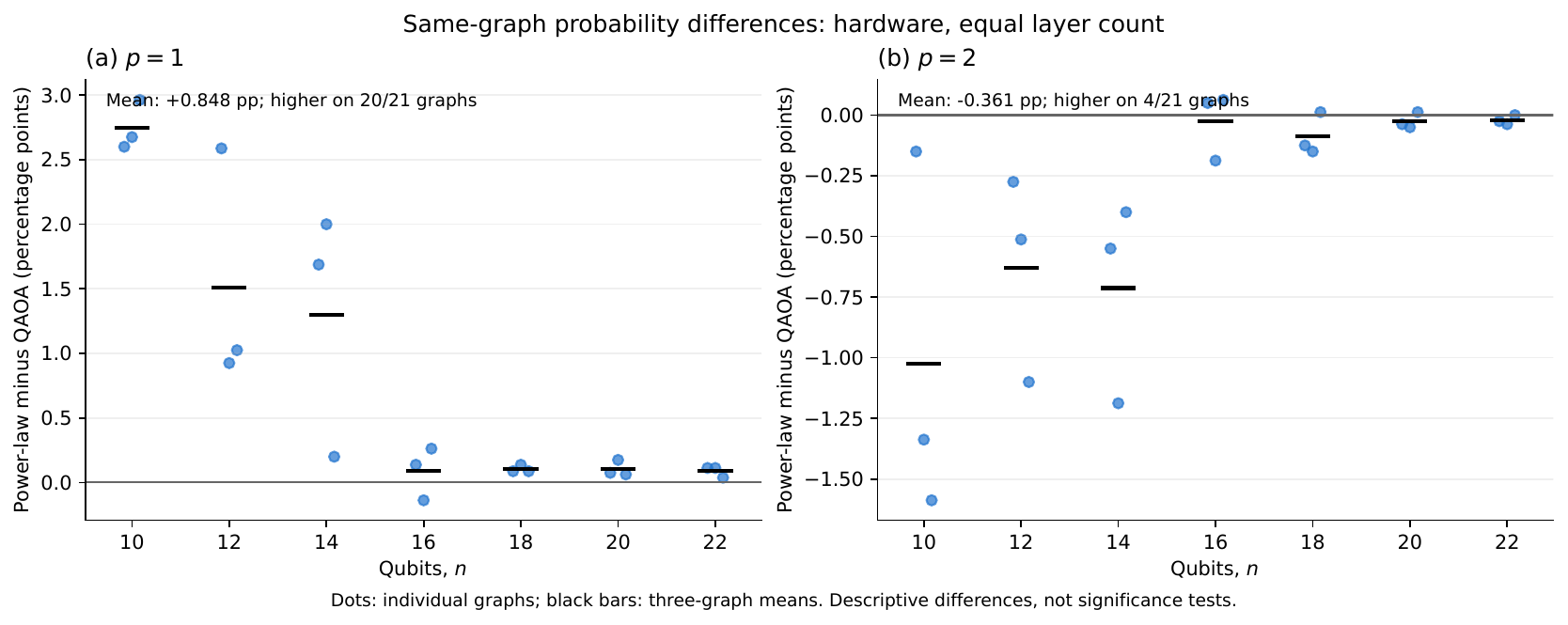}
\caption{Raw hardware paired probability differences at (a) $p=1$ and (b) $p=2$. Blue dots show the three individual graphs at each size, with small horizontal offsets for visibility; black bars show their means. Positive values favor power-law DC-QAOA. All 21 graphs are retained at each depth. }
\label{fig:hardware-paired}
\end{figure}
\begin{figure}[H]
\centering

\includegraphics[width=0.98\linewidth]{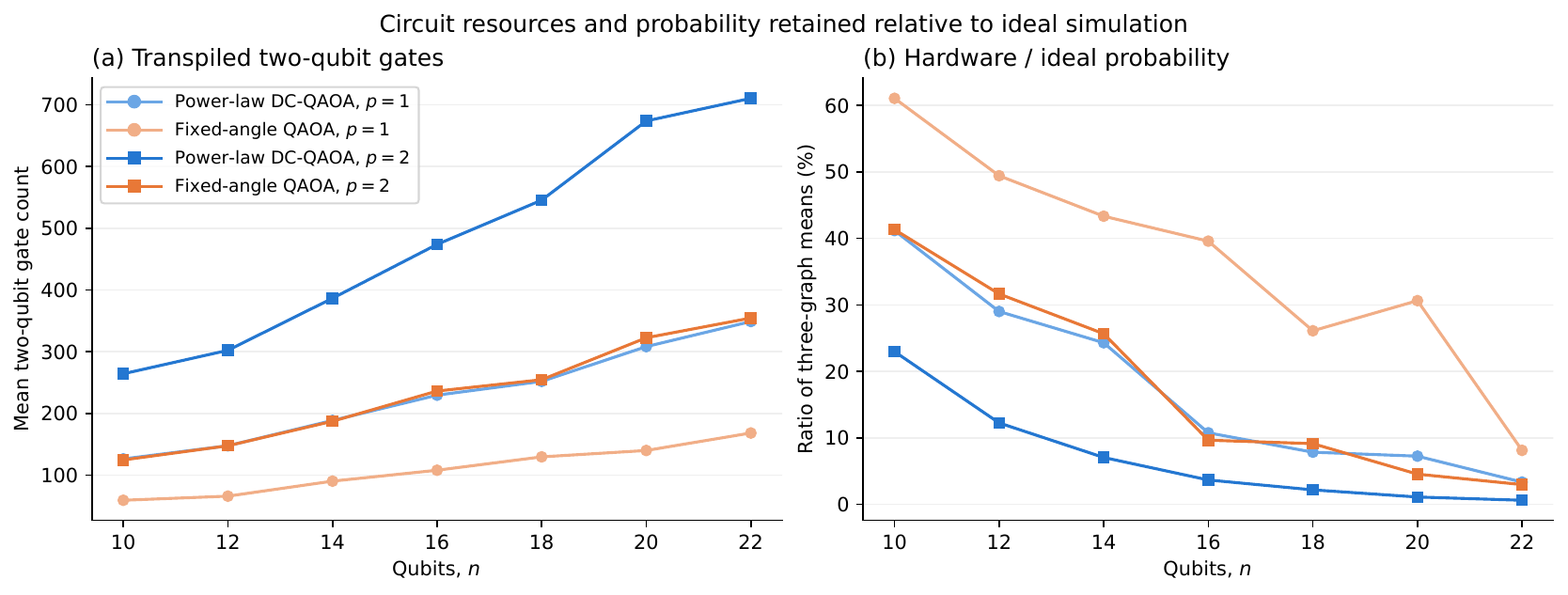}
\caption{Circuit resources and raw hardware probabilities relative to ideal references. (a) Mean transpiled two-qubit gate counts. (b) The ratio of the mean hardware probability to the mean ideal probability, expressed as a percentage. Each point uses three graphs. Blue denotes power-law DC-QAOA and orange fixed-angle QAOA; light circles show $p=1$ and darker squares $p=2$.}
\label{fig:hardware-resources}
\end{figure}
\endgroup

\clearpage
\section*{S4. Graph ensembles and edge-weight distributions}

Tables S2 and S3 collect the graph-generation and edge-weight sampling definitions used in the main-text comparisons. A zero-edge draw is replaced by a cycle. Edge weights are sampled independently. The probability comparisons use a balanced selection of three graphs per size--family--weight cell; the selection procedure is specified in Section~S8. These sampling choices do not alter the graph-generation rules.
{
\begin{table}[htbp]

\centering
\caption{Graph-family definitions. The regular3--regular6 row represents four families, giving ten families in total}
\small\renewcommand{\arraystretch}{1.35}
\begin{tabular}{p{0.22\linewidth}p{0.7\linewidth}}\hline
Family & Generation rule \\ \hline
\texttt{tree} & Random labeled tree. \\
\texttt{ring} & Cycle. \\
\texttt{regular3}--\texttt{regular6} & Random $d$-regular graphs with $d=3$--6. \\
\texttt{er3} & $G(n,3/(n-1))$. \\
\texttt{er05} & $G(n,0.5)$. \\
\texttt{community} & Two-block stochastic block model with block sizes $\lfloor n/2\rfloor$ and $\lceil n/2\rceil$, within-block probability 0.45 and between-block probability 0.08. \\
\texttt{bipartite} & Random bipartite graph with block sizes $\lfloor n/2\rfloor$ and $\lceil n/2\rceil$ and cross-block probability 0.35. \\ \hline
\end{tabular}
\end{table}
\begin{table}[htbp]

\centering
\caption{Positive edge-weight distributions. Lognormal parameters specify the mean and standard deviation of the underlying normal distribution}
\small\renewcommand{\arraystretch}{1.35}
\begin{tabular}{p{0.18\linewidth}p{0.74\linewidth}}\hline
Distribution & Sampling rule \\ \hline
Unit & $J=1$. \\
Uniform & $J\sim U(0.5,1.5)$. \\
Bimodal & $J\in\{0.5,1.5\}$ with equal probability. \\
Lognormal & $J=\operatorname{clip}(\operatorname{LogNormal}(0,0.65),0.2,5)$. \\
Heavy tail & $J=\operatorname{clip}(\operatorname{LogNormal}(0,1.2),0.2,10)$. \\ \hline
\end{tabular}
\end{table}
}
\clearpage
\section*{S5. Sensitivity to low-degree extrapolation}

The power-law coefficients are fitted using $D>1$. We examine extrapolation on the 11 graphs with $D\leq1$ in the test pool, at $p=1,2,3,4,6$. Their mean degrees are $2/3$, $0.8$, or $1$. We use the same frozen coefficients without a low-degree fallback or refitting. One graph is shared with the 1050-graph sample and its 600-graph depth-six subset; the other ten are diagnostic samples and do not enter those averages.

\begin{figure}[H]
\centering
\includegraphics[width=0.95\linewidth]{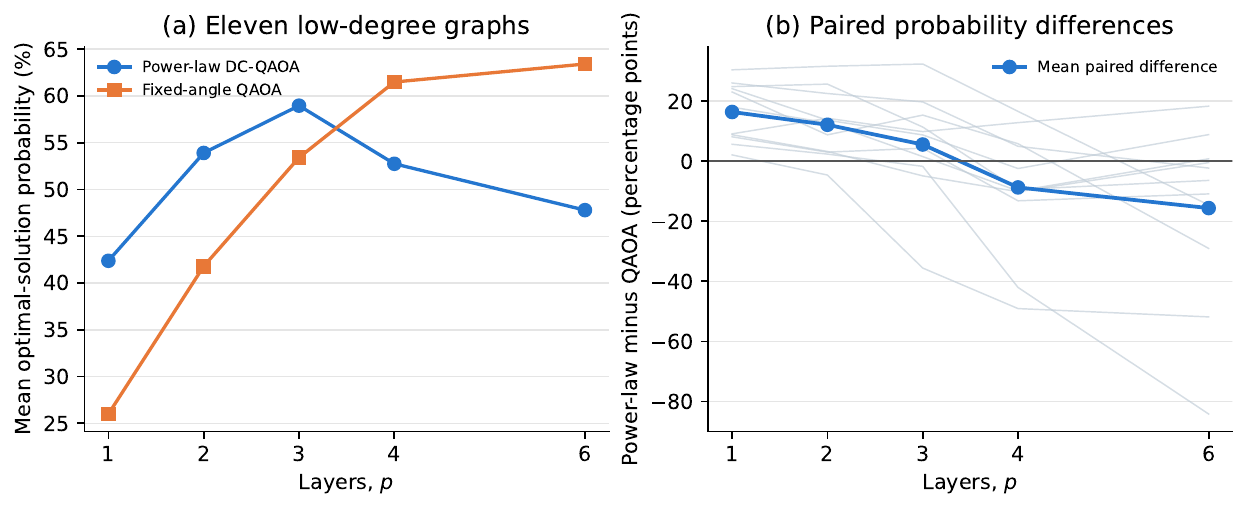}
\caption{Low-degree extrapolation on the same 11 graphs at each depth. (a) Mean optimal-solution probabilities. (b) Paired differences, $100(P_{\rm PL}-P_Q)$, in percentage points; thin gray lines show individual graphs and the blue line their mean. All 55 graph--depth pairs use PennyLane exact statevectors. No confidence intervals are assigned to this small diagnostic set.}\label{fig:low-degree-probability}
\end{figure}

Fig.~\ref{fig:low-degree-probability} compares power-law DC-QAOA with fixed-angle QAOA using the optimal-solution probability defined in the main text. The mean paired differences are $+16.38$, $+12.12$, and $+5.55$ percentage points at $p=1,2,3$, but $-8.74$ and $-15.62$ at $p=4,6$. Thus, a single formula for all positive degrees does not ensure a probability advantage below its fitting range. These 11 graphs form a small diagnostic set rather than a representative sample of low-degree graphs.

\section*{S6. Protocol for formula-initialized refinement} This protocol applies to the 12 held-out graphs at \(n=12\) and \(p=2\) in the main-text refinement experiment, with one formula-initialized and 20 randomly initialized L-BFGS-B runs per graph.

We use the PennyLane \cite{bergholm2018} \texttt{lightning.qubit} backend with \texttt{complex128} statevectors and analytic adjoint gradients, with \texttt{maxiter} = 200, \texttt{ftol} = $10^{-10}$, and \texttt{gtol} = $10^{-5}$. 

The objective is the Ising energy. One call returns both its value and its adjoint gradient, unlike the energy-only evaluation count in Section~\mbox  {III-C} of the main text. The reported trajectories and terminal ratios are best-so-far approximation ratios over recorded evaluations, including line-search trials; they need not equal the ratio at the optimizer-returned angles. The 99\% gain point is the first call attaining 99\% of that run's observed improvement over its initial ratio. Before optimization, we cross-check each power-law formula energy against an independent Qulacs evaluation to within $10^{-7}$.

For each layer, the random initializations are drawn independently as $\gamma_l\sim U(-1.2,0.2)f_D(D)/s_J$, $\beta_l\sim U(-0.1,0.9)$, and $\alpha_l\sim U(-0.4,0.2)/s_J$. Here $f_D(D)=\arctan(1/\sqrt{D-1})$ for the $D>1$ graphs used in this experiment. This factor specifies the random-start distribution; the deterministic initialization uses the power-law rule. Both initialization strategies use the bounds $|\gamma_l|\leq3\pi$, $|\beta_l|\leq\pi$, and $|\alpha_l|\leq2\pi$. The random seed is 20261024. This experiment uses a different initialization distribution and gradient protocol from the comparison in Section~\mbox  {III-C} of the main text.

\section*{S7. Protocol for the random-start optimizer benchmark}
This protocol applies to the three-optimizer comparison in Section~\mbox  {III-C} of the main text, rather than the refinement experiment in Section~S6. It uses 12 held-out graphs at \(n=12\) and \(p=2\), with three graphs for each combination of 3-regular or ER3 structure and unit or uniform weights. The run manifest lists the selected graph IDs. Each optimizer uses the same 20 initial angle vectors on each graph, giving 240 runs per optimizer and 720 runs in total.

For each layer, the initial angles are sampled independently from \(\gamma_l,\alpha_l\sim U(-\pi,\pi)\) and \(\beta_l\sim U(-\pi/2,\pi/2)\), without graph-dependent rescaling. The optimization bounds are the same intervals. We use NumPy's \texttt{default\_rng}; each seed is the big-endian integer formed from the first four SHA-256 digest bytes of the string \texttt{20261022:fingerprint:start}, with the graph fingerprint and zero-based start index substituted. Each run record stores the resulting seed and the full initial vector in the order \((\gamma_1,\gamma_2,\beta_1,\beta_2,\alpha_1,\alpha_2)\).

Circuits are evaluated with PennyLane 0.45.0 and \texttt{lightning.qubit}, using \texttt{shots=None} to compute exact expectation values. The QNode uses the NumPy interface with \texttt{diff\_method=None}. The power-law formula reference is evaluated separately for each graph and cross-checked with Qulacs. These reference evaluations are separate from each optimizer's search budget.

L-BFGS-B uses SciPy's \texttt{minimize} with no supplied Jacobian, \texttt{maxiter}=500, and \texttt{ftol}=\(10^{-10}\); gradients are estimated numerically by SciPy, and the remaining L-BFGS-B options are not overridden. COBYLA uses the same bounds with \texttt{maxiter}=500, \texttt{rhobeg}=0.5, and \texttt{tol}=\(10^{-4}\). Adam uses one-sided finite differences with step \(10^{-4}\), switching to a backward difference if a forward probe would exceed the upper bound. Its learning rate is 0.08, its two moment coefficients are 0.9 and 0.999, and its denominator offset is \(10^{-8}\). The moments are bias-corrected, and each update is clipped to the parameter bounds. Adam has no separate convergence stopping criterion.

An objective wrapper enforces a limit of 500 energy evaluations per run, including all finite-difference and line-search calls. Runs may stop earlier under the optimizer's own termination criteria. The best-so-far approximation ratio includes every recorded evaluation, including gradient probes; its last value is carried forward after early termination when forming the mean trajectory. 

A run attains the reference when its best-so-far ratio reaches the same graph's formula-based ratio minus 0.001 within the evaluation budget.

\clearpage
\begingroup
\setlength{\parskip}{3pt}
\section*{S8. Offline training, frozen comparators, and\\reproducibility records}

The graph data comprise two cohorts, each divided into training, validation, and test sets. The 10--16-qubit cohort contains 4000 training, 2000 validation, and 4000 test graphs; the 18--22-qubit cohort contains 3000 training, 1500 validation, and 3000 test graphs. The combined test pool contains 7000 graphs, with 20 graphs in each of the 350 size--family--weight cells. Graph-generation and edge-weight sampling rules are given in Section~S4.

For the probability evaluation, we use a fixed, balanced sample of 1050 graphs from this pool to limit statevector evaluation cost while maintaining equal representation of the size--family--weight cells. We exclude the 200 graphs used in the earlier exploratory pilot, then select three eligible graphs per cell. Within each cell, eligible graph IDs are ordered by the hexadecimal SHA-256 digest of the UTF-8 string formed by the seed \texttt{20261027}, a vertical bar, and the graph ID; the first three are selected. This ordering uses graph IDs, not performance scores. The resulting selection was frozen before the present probability evaluation and reused across both methods and all four main depths.

The selected graphs span $n=10,12,14,16,18,20,22$, ten graph families, and five weight distributions, with 150 graphs per size. The depth-six extension uses the same 600 selected graphs with $n=10,12,14,16$; no new selection is made at $p=6$. The 11-graph low-degree diagnostic is treated separately, with one graph shared with the selected sample and its depth-six extension. The other ten diagnostic graphs do not enter the 1050- or 600-graph averages.

The offline training runs use PennyLane \texttt{lightning.gpu} with \texttt{complex64} statevectors and analytic adjoint gradients. Each graph--depth pair has eight L-BFGS-B starts, with \texttt{maxiter}=200, \texttt{maxfun}=850, \texttt{ftol}=\(10^{-6}\), and \texttt{gtol}=\(10^{-3}\). The bounds are \(|\gamma_l|\leq3\pi\), \(|\beta_l|\leq\pi\), and \(|\alpha_l|\leq2\pi\). Final energies are recomputed with \texttt{complex128} statevectors before selecting the retained angle vector.

The starts combine structured schedules and independent random draws. QAOA uses a linear ramp and, when available, interpolation from the best schedule at the preceding evaluated depth; the remaining starts are random. DC-QAOA first uses the best same-depth QAOA angles with all counterdiabatic angles set to zero, then an interpolated schedule or, at the first depth, a ramp; the remaining six starts are random. Random starts draw \(\gamma_l\) from \(U(-1.2,0.2)f_D(D)/s_J\), \(\beta_l\) from \(U(-0.1,0.9)\), and \(\alpha_l\) from \(U(-0.4,0.2)/s_J\). The exact ramp, interpolation, and seed construction are implemented in \texttt{optimization.py} and \texttt{eqdepth.py}; the run configuration uses base seed 20261013. Restart records retain the initial angles and optimization seeds.

For historical random starts and QAOA comparators, $f_D(D)$ equals $\arctan(1/\sqrt{D-1})$ for $D>1$ and $1$ otherwise. It specifies those protocols, not the current DC-QAOA angle-assignment rule.

The probability results in main-text Figs.~3, 7, and 8 and in Sections~S5, S9, and S10 use PennyLane \texttt{lightning.qubit}, \texttt{complex128} statevectors, and \texttt{shots=None}, with the cost--ordered CD--mixer sequence. The main probability campaign records PennyLane 0.45.0. Each paired record is checked against its graph identity, frozen angle assignment, and exact Ising extrema. These evaluations are separate from the offline training and the optimization protocols in Sections~S6 and S7.

For each graph, we enumerate the diagonal Ising energies and sum the final-state probabilities over all minimizing bit strings, including spin-flipped partners. With $W=\sum_eJ_e$, a state belongs to the numerical optimum set when $|E(z)-E_{\min}|\leq10^{-10}\max(1,W)$. The implementation checks that using $10^{-8}\max(1,W)$ selects exactly the same states; an ambiguous mask is rejected. State normalization is checked to within $10^{-9}$. Thus, the reported probabilities are statevector sums rather than finite-shot estimates.

Probability confidence intervals use 2000 graph-bootstrap resamples (seed 20261027) and the 2.5th and 97.5th percentiles. Pooled intervals preserve the sample count within each size stratum; size-specific intervals resample graphs within that size. Both methods remain paired on each graph. For cross-depth summaries, the same sampled graph carries all its depths, so averaging over depths does not increase the number of independent graphs. Intervals are pointwise, without multiple-comparison correction, and describe graph-sample variation rather than measurement shot noise. The heatmaps and the 11-graph low-degree diagnostic report descriptive means.

The reproducibility records retain graph IDs and fingerprints, the frozen model and QAOA choices, assigned angles, simulator configuration, and probabilities. The main paired table contains 4200 graph--depth records; the same-600-graph table contains 3000 across $p=1,2,3,4,6$, reusing its first four depths. The low-degree table contains 55 records, five shared with these samples. After removing overlaps, these probability evaluations cover 1060 graphs and 4850 distinct graph--depth pairs. Historical full-pool Qulacs energy results remain separate from the PennyLane probability records.

Table~S4 records the nine archived cohort-specific QAOA choices.

\begin{table}[!htbp]
\centering
\caption{Validation-selected fixed-angle QAOA comparators. The archived method keys are \texttt{qaoa\_trained} for the cohort-trained medoid and \texttt{qaoa\_csv} for the transferred table with RMS-weight and degree scaling}
\label{tab:qaoa-comparators}
\begingroup
\begin{tabular}{lll}\hline
Qubit cohort & Depth & Selected comparator \\\hline
10--16 & 1 & Cohort-trained medoid \\
10--16 & 2 & Cohort-trained medoid \\
10--16 & 3 & Cohort-trained medoid \\
10--16 & 4 & Cohort-trained medoid \\
10--16 & 6 & Transferred table \\
18--22 & 1 & Cohort-trained medoid \\
18--22 & 2 & Cohort-trained medoid \\
18--22 & 3 & Cohort-trained medoid \\
18--22 & 4 & Transferred table \\\hline
\end{tabular}
\endgroup
\end{table}

The archived training environments record Python 3.12.15, NumPy 2.5.3, SciPy 1.18.1, PennyLane 0.45.1, and PennyLane-Lightning-GPU 0.45.0. These versions describe the offline training runs, not the separate optimizer benchmark in Section~S7. For the exact MaxCut reference, the CPU code enumerates \(2^{n-1}\) spin configurations using global spin-flip symmetry and records the minimum and maximum Ising energies. The resulting extrema are associated with each graph ID and fingerprint.

Table~1 of the main text lists five coefficients per layer: $A_{\gamma,l,p}$, $b_{\gamma,l,p}$, $B_{l,p}$, $A_{\alpha,l,p}$, and $b_{\alpha,l,p}$. The 16 layers across $p=1,2,3,4,6$ require 80 coefficients plus $D_0=38/11$. Evaluations use full-precision values. The amplitudes and exponents are fitted separately to the sign-aligned training angles using the degree-bin regression in Section~II-C; mixer values are retained from the pooled offline schedule.

The eight entries at $p=1$--4 provide the comparators for the 4200 main graph--depth pairs, and the smaller-cohort $p=6$ entry provides the transferred-angle comparator for the 600 additional pairs in Section~S9. These choices minimize the mean normalized Ising energy on the validation graphs and were not reselected to maximize optimal-solution probability. Each cohort's \texttt{FROZEN\_MODEL.json} stores the selected method, candidate validation scores, and trained QAOA coefficients. The transferred candidate reads the \(q=2\) rows of \texttt{original\_transfer\_data.csv}; it multiplies the stored cost angles by \(-f_D(D)/s_J\) and leaves the mixer angles unchanged. The RMS-only candidate was not selected in any of the nine comparisons.

\endgroup
\clearpage
\begingroup
\section*{S9. Depth-six evaluation}

We compare both methods on the same 600 graphs with $n=10,12,14,16$ at $p=1,2,3,4,6$, with 150 graphs per size. The first four depths reuse the corresponding main-sample results; the depth-six probabilities use the same PennyLane protocol in Section~S8. The coefficients at $p=6$ were fitted offline, so this is an evaluation at a trained depth, without test-instance refitting.

The depth-six QAOA comparator is the validation-selected transferred-angle schedule in Table~\ref{tab:qaoa-comparators}; the first four depths use cohort-trained medoids. Cross-depth comparisons therefore assess the frozen baseline strategy at each depth, rather than truncations of one QAOA schedule.

Table~\ref{tab:p6-probability} and Fig.~\ref{fig:p6-probability} report the probability comparison on this fixed sample. At $p=6$, the mean probabilities are 36.810\% and 26.691\%, with a mean paired difference of 10.119 percentage points and a pointwise 95\% interval of $[9.206,11.028]$. Both mean probabilities and the mean paired probability difference increase across the displayed depths on these 600 graphs. Because this sample contains only $n\leq16$, its $p=6$ values should be compared with the $p=1$--4 rows of this table rather than with the 1050-graph means.

\begin{table}[H]
\centering\small
\caption{Optimal-solution probabilities on the same 600 graphs at five depths. Probabilities are percentages; differences and their pointwise 95\% paired-bootstrap intervals are percentage points}\label{tab:p6-probability}
{
\begin{tabular}{crrrl}\hline
$p$ & $P_{\rm PL}$ (\%) & $P_Q$ (\%) & $100\Delta P$ & 95\% interval \\ \hline
1 & 7.303 & 2.357 & +4.945 & $[4.586, 5.300]$ \\
2 & 12.445 & 6.940 & +5.506 & $[5.177, 5.836]$ \\
3 & 20.005 & 12.756 & +7.249 & $[6.811, 7.673]$ \\
4 & 26.480 & 18.885 & +7.595 & $[7.108, 8.080]$ \\
6 & 36.810 & 26.691 & +10.119 & $[9.206, 11.028]$ \\
\hline\end{tabular}}
\end{table}

\begin{figure}[H]
\centering
\includegraphics[width=0.95\linewidth]{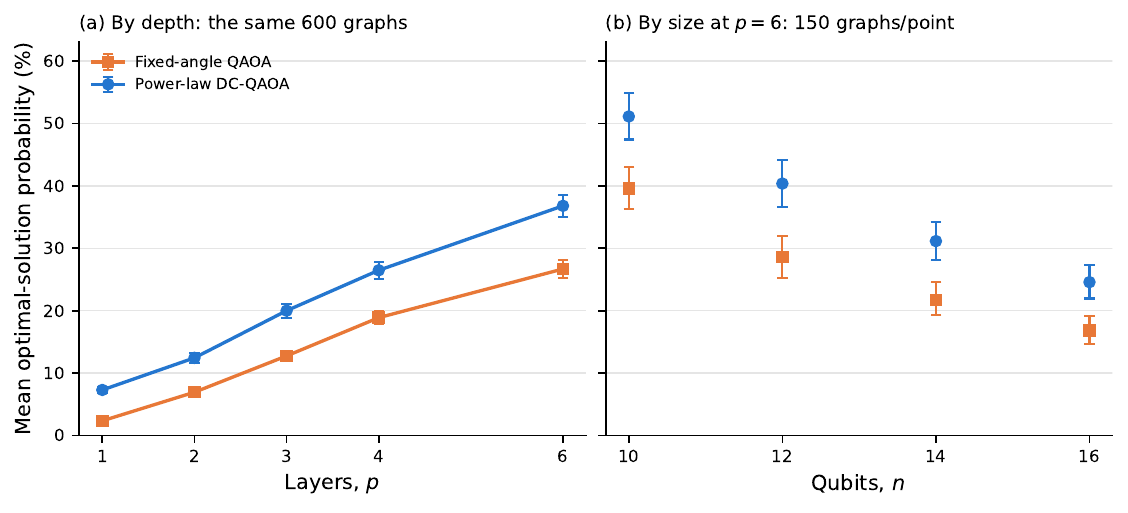}
\caption{Depth-six extension. (a) Mean optimal-solution probabilities on the same 600 graphs at every displayed depth. (b) Size-resolved results at $p=6$, with 150 graphs per point. Error bars are pointwise 95\% graph-bootstrap intervals from 2000 resamples. Pooled resampling is stratified by size and carries both methods and all depths together; size-specific intervals resample within each size.}\label{fig:p6-probability}
\end{figure}
\endgroup

\clearpage
\section*{S10. Size-resolved comparisons, aggregation checks, and approximation ratios}
\begin{figure}[H]
\centering
\includegraphics[width=0.95\linewidth]{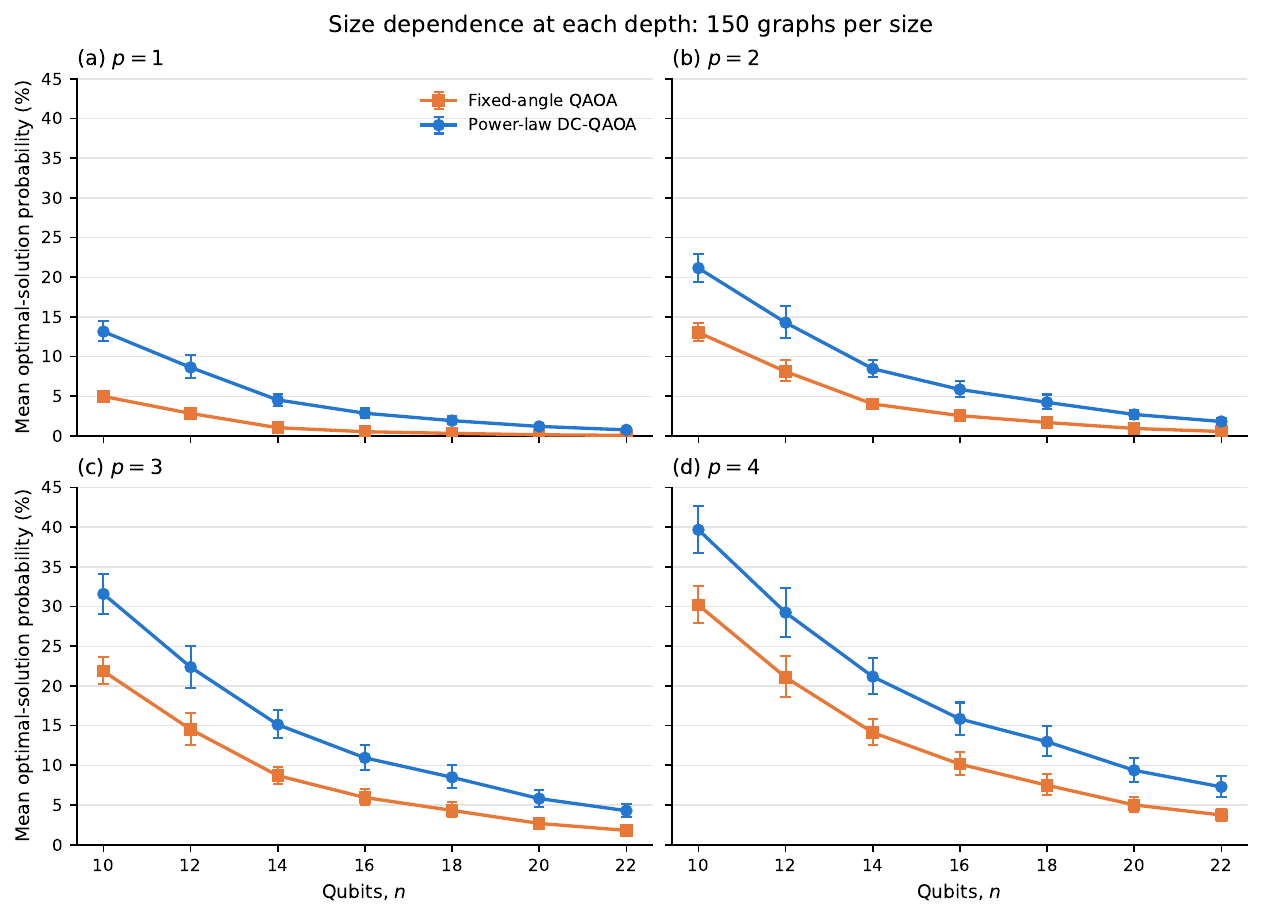}
\caption{Optimal-solution probabilities by graph size at each depth: (a)--(d) show $p=1,2,3,4$, respectively. Each point uses the same 150 graphs per size, balanced across ten graph families and five weight distributions. Error bars show pointwise 95\% intervals from 2000 graph-bootstrap resamples within each size, retaining method pairing. Lines guide the eye.}\label{fig:size-by-depth}
\end{figure}

Fig.~\ref{fig:size-by-depth} separates the four depths averaged in main-text Fig.~3(b). Each point uses 150 graphs at a fixed size and depth. The main-text average first gives each of the four depths equal weight within a graph, then averages those graph-level values; its bootstrap draws carry all four depths together. Thus, each size still contributes 150 independent graphs, not 600.

Main-text Fig.~7(b) applies the same graph-level depth averaging within each family--size cell, using 15 graphs. Although all 70 averaged cells are positive, one of the 280 family--size--depth cells has a negative mean: bipartite graphs at $n=22$, $p=4$, with $-0.048$ percentage points (7 negative and 8 positive individual differences). For that family and size, the means at $p=1,2,3,4$ are $+1.189$, $+1.509$, $+1.556$, and $-0.048$ percentage points, giving $+1.052$ after equal-depth averaging. Fig.~\ref{fig:size-by-depth} pools graph families and therefore does not display that family-specific exception separately.

Table~\ref{tab:ratio-1050} reports the mean approximation ratio $r=\mathbb{E}[C]/C_{\max}$ for the same 4200 ideal-circuit records. Power-law DC-QAOA has the higher mean ratio at every depth, but the mean ratio difference decreases from 0.0659 at $p=1$ to 0.0111 at $p=4$, whereas the mean probability difference increases from 3.30 to 6.25 percentage points. The number of graphs with a higher ratio for power-law DC-QAOA falls from 1047 to 888 of 1050 over the same depths.

\begin{table}[H]
\centering\small
\caption{Mean approximation ratios on the 1050-graph sample. $\Delta r$ is the mean paired difference $r_{\rm PL}-r_Q$; the last column counts graphs with $r_{\rm PL}>r_Q$}\label{tab:ratio-1050}
\begin{tabular}{ccccc}\hline
$p$ & $r_{\rm PL}$ & $r_Q$ & $\Delta r$ & $r_{\rm PL}>r_Q$ \\ \hline
1 & 0.8383 & 0.7723 & 0.0659 & 1047 \\
2 & 0.8704 & 0.8386 & 0.0318 & 1006 \\
3 & 0.8949 & 0.8749 & 0.0200 & 953 \\
4 & 0.9095 & 0.8984 & 0.0111 & 888 \\ \hline
\end{tabular}
\end{table}

%% file: abstract.tex
\begin{abstract}
Optimizing parameters separately for each problem instance limits the reuse of variational quantum algorithms. We develop empirical scaling laws for parameter setting in the digitized counterdiabatic quantum approximate optimization algorithm (DC-QAOA) for weighted MaxCut. The prescription combines exact covariance under uniform edge-weight rescaling with power laws in mean degree fitted separately to offline-optimized cost and counterdiabatic angles. After the weight and degree scales are removed, the optimized angles of different training graphs cluster around common amplitudes, so a small coefficient table suffices. Frozen coefficients assign angles from a target graph's mean degree and root-mean-square edge weight without target-instance variational optimization.
  In ideal-circuit simulations across 1050 test graphs spanning ten graph families and five positive edge-weight distributions, mean optimal-solution probability is higher than that of the fixed-angle quantum approximate optimization algorithm (QAOA) at each depth from one to four layers, reaching 19.36\% versus 13.11\% at four layers. A six-layer extension on 600 graphs also shows a mean probability advantage. Hardware execution on the IBM Quantum processor ibm\_fez for 21 unit-weight 3-regular graphs shows a mean probability advantage at one layer, while the ordering reverses at two layers. In separate energy-based experiments, direct assignment gives a higher mean approximation ratio than random-start optimization reaches within 500 evaluations, and refinement started from the formula angles improves it further. These results show how a scaling-based representation of optimized angles can guide parameter reuse across graph instances, enabling direct parameter assignment without target-instance variational optimization and providing initial angles for local refinement.
\end{abstract}

%% file: body.tex
\section{Introduction}\label{introduction}

Combinatorial optimization is central to decisions in transportation, production, and financial infrastructure. Vehicle routing, job-shop scheduling, and transaction settlement have each motivated investigations of quantum optimization methods \cite{azad2022,amaro2022,braine2021}. Applying these algorithms to new instances requires considering both solution quality and parameter-setting costs.

The quantum approximate optimization algorithm (QAOA) is a widely studied framework for quantum optimization \cite{farhi2014}. Its circuit alternates a \emph{cost operator}, which encodes the optimization objective, with a \emph{mixer operator}, which enables transitions among candidate solutions. The resulting \(p\)-layer circuit has variational angles that a classical optimizer typically selects through repeated evaluations of the quantum objective. Structured angle patterns and interpolation-based initialization can make this search more effective \cite{zhou2020}. Random initialization can also lead to suboptimal local minima in studied QAOA landscapes, motivating more deliberate initialization schemes \cite{sack2021}. These strategies improve angle search, but avoiding that search on each new graph remains a separate objective.

 Counterdiabatic driving motivates additional operations in the QAOA ansatz. In an ideal continuous evolution, a counterdiabatic correction can suppress excitations caused by a finite-rate change of Hamiltonian \cite{berry2009}. Approximate local corrections \cite{sels2017} and nested-commutator constructions \cite{claeys2019} address the difficulty of implementing exact counterdiabatic terms in many-body systems. Digitized counterdiabatic quantum optimization has also been studied as a gate-based approach for Ising models \cite{hegade2022}; its protocol differs from the variational circuit considered here. The connection between counterdiabaticity and QAOA has been examined theoretically \cite{wurtzlove2022}, and digitized counterdiabatic QAOA (DC-QAOA) incorporates counterdiabatic-inspired operations into a gate-based alternating circuit \cite{chandarana2022}. The resulting ansatz can achieve better performance than standard QAOA at a fixed number of layers in previously studied settings \cite{chandarana2022}. Subsequent studies have examined the interplay between counterdiabatic corrections, circuit depth, and parameter count \cite{vizzuso2024}, as well as the optimization landscapes and classical optimizers of related digitized-counterdiabatic circuits \cite{xu2024}.

The counterdiabatic extension also creates a parameter-setting problem. In the explicit first-order circuit studied here, each layer has the usual QAOA cost and mixer angles, \(\gamma_l\) and \(\beta_l\), together with a counterdiabatic angle \(\alpha_l\) controlling two-qubit rotations associated with \(Y_iZ_j\) and \(Z_iY_j\) terms. These operations increase the number of entangling gates per layer. When the method is reused across graphs, optimizing the enlarged parameter vector separately for each target instance requires additional objective-function evaluations. Repeated angle searches can therefore limit the practical benefit of improved solution quality at fixed depth. We ask how edge-weight and degree scales can guide parameter transfer in DC-QAOA and whether the resulting prescription can replace variational angle search on a new graph.

\begin{figure*}[t]
\centering
\includegraphics[width=0.95\linewidth]{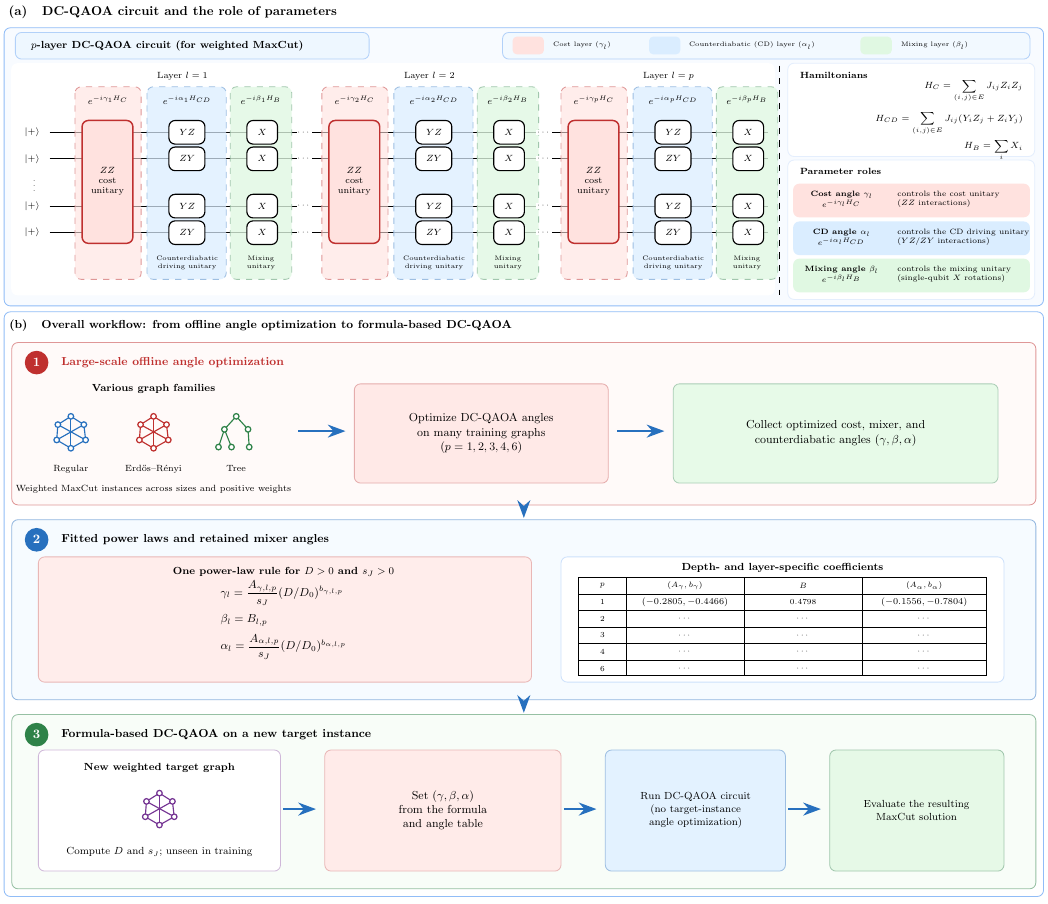}
\caption{
Circuit and parameter-setting overview. (a) Each DC-QAOA layer applies an edge-weighted cost block, an ordered counterdiabatic block, and a mixing block controlled by $\gamma_l$, $\alpha_l$, and $\beta_l$, respectively. (b) Offline-optimized training angles determine depth- and layer-specific power-law amplitudes and exponents for the cost and counterdiabatic angles; mixer angles are retained from the offline schedule. The coefficients are fitted using training records with $D>1$. The displayed $p=1$ coefficients are rounded for presentation. Exact gate ordering and angle conventions are given in Section~\ref{weighted-maxcut-circuit-and-scaling-analysis}.}\label{fig:1}
\end{figure*}

Prior work has explored several approaches to transferring and predicting QAOA parameters. For standard QAOA, the objective value at fixed parameters can concentrate across typical instances \cite{brandao2018}. Separately, studies of optimized parameters have examined their concentration analytically at low depth and numerically at higher depth \cite{akshay2021}. Fixed-angle schedules have been investigated for regular MaxCut graphs \cite{wurtzlykov2021}. An infinite-size analysis has been used to predict QAOA angles for finite random graphs \cite{boulebnane2021}, while topology-based classical inference \cite{streif2020} and offline-trained neural predictors \cite{amosy2024} can reduce or, in their studied settings, eliminate iterative angle search on a new instance. Cross-graph transfer has also been related to local graph structure and similarity between donor and target graphs \cite{galda2021,galda2023}. A broader study of graph families and weight distributions finds that the success of QAOA initialization strategies depends on instance characteristics \cite{katial2025}.

For weighted MaxCut, parameter rescaling adapts previously obtained QAOA angles to new weight distributions \cite{shaydulin2023,sureshbabu2024}. In particular, Sureshbabu \emph{et al.} \cite{sureshbabu2024} use root-mean-square (RMS) edge-weight normalization, leave mixer angles unscaled, and discuss a finite-degree correction for the cost angles. We retain RMS-weight normalization and unscaled mixer angles, and fit separate degree power laws to the cost and counterdiabatic angles obtained from offline DC-QAOA optimization.

Most of these parameter-transfer approaches concern conventional QAOA. For DC-QAOA, meta-learning has been used to obtain better initial parameters for subsequent variational optimization \cite{chandarana2023}. Liu and Wang studied parameter concentration in a nested-commutator DC-QAOA ansatz and introduced instance-sequential training, which initializes optimization on larger graphs from trained smaller subgraphs \cite{liuwang2025}. They compare circuits at equal CNOT counts and examine how the results vary with system size. These DC-QAOA studies investigate more efficient \emph{training} and, in the latter case, resource-matched ansatz performance. Our goal is a compact, frozen prescription that assigns all three DC-QAOA angle families to a new weighted graph without variational angle optimization on that graph.

For positive-weight MaxCut, we construct a formula-based DC-QAOA prescription from extensive \emph{offline} optimization (Fig.~\ref{fig:1}(b)). A depth-indexed table stores layer-specific amplitudes and degree exponents for the cost and counterdiabatic angles, together with mixer angles and a shared reference degree. For a new graph, its mean degree $D$ and RMS edge weight $s_J$ determine the assigned angles through the frozen power-law rule. The implemented circuit is exactly covariant under uniform edge-weight rescaling. Generator second moments provide an inverse-square-root degree reference scale, while the degree exponents are fitted to training angles. Removing these weight and degree scales reveals concentration around common, layer-dependent amplitudes. The same rule is defined for every $D>0$ and $s_J>0$; applying it at $D\leq1$ extrapolates beyond the fitting range. Angle assignment requires no target-instance quantum-objective calls.

 In ideal-circuit simulations, we evaluate the frozen prescription using optimal-solution probability on a balanced sample of 1050 graphs spanning ten graph families, five positive-weight distributions, and 10--22 qubits at depths $p=1$--4. The depth-six extension uses the same 600 graphs with 10--16 qubits from this sample. We compare against validation-selected fixed-angle QAOA at the same number of layers, without optimizing either method on the target graph. Power-law DC-QAOA yields higher mean optimal-solution probabilities at all tested depths in these samples, with variation across graph instances. A separate experiment on 12 graphs measures the cost of random-start energy optimization and evaluates formula-initialized refinement through approximation-ratio trajectories. Supplementary diagnostics examine low-degree extrapolation and compare probabilities separately by size and depth. We also test the frozen prescription on hardware using unit-weight 3-regular graphs.

This work makes three contributions:

\begin{enumerate}
\def\labelenumi{\arabic{enumi}.}
\tightlist
\item

\textbf{Parameter scaling laws.} We combine exact covariance under uniform edge-weight rescaling with empirical, depth- and layer-specific degree power laws for the cost and counterdiabatic angles.
\item
    
\textbf{A compact representation and explicit assignment rule.} In the rescaled coordinates, optimized angles from different training graphs cluster around common amplitudes, so a frozen formula can assign circuit angles from graph statistics alone, without target-instance variational optimization.
\item

\textbf{Evaluation of direct assignment and refinement.} We assess the rule through optimal-solution probabilities relative to fixed-angle QAOA and separate energy-based optimization experiments, including the cases where it loses, notably low-degree graphs at higher depth.
\end{enumerate}

The remainder of this paper is organized as follows. Section~\ref{methods-from-offline-optimized-angles-to-a-formula-based-prescription} presents the angle prescription, implemented circuit, and offline construction and evaluation protocol. Section~\ref{results-performance-and-limits-of-scaling-guided-parameter-setting} reports optimal-solution probabilities from ideal-circuit simulations and hardware execution, followed by the cost of random-start energy optimization, formula-initialized refinement, and analyses of graph structure and edge weights. Section~\ref{discussion} discusses the scope and limitations of the findings, and Section~\ref{conclusion} concludes. Appendix A proves the uniform edge-weight scaling relation and the resulting invariance of optimal-solution probability, and Appendix B derives the generator second-moment identities and their connection to the empirical power-law rule. The Supplementary Material provides additional scaling diagnostics, graph-generation and weight-sampling definitions, low-degree and depth-six probability results, size-resolved comparisons, and the optimization and reproducibility protocols.

\section{Methods: From offline optimized angles to a formula-based prescription}\label{methods-from-offline-optimized-angles-to-a-formula-based-prescription}

\subsection{Formula-based angle setting with graph-dependent rescaling}\label{formula-based-angle-setting-with-graph-dependent-rescaling}

For a MaxCut graph \(G=(V,E)\) with \(n\) vertices, \(|E|>0\), and positive edge weights \(J_{ij}\), let \(\gamma_l\), \(\beta_l\), and \(\alpha_l\) denote the cost, mixer, and counterdiabatic angles at layer \(l\) of a depth-\(p\) circuit. We assign the angles using a depth- and layer-specific power-law prescription:
\begin{equation}
\boxed{\begin{aligned}
\gamma_l(G)&=\frac{A_{\gamma,l,p}}{s_J(G)}\left(\frac{D(G)}{D_0}\right)^{b_{\gamma,l,p}},\\
\beta_l(G)&=B_{l,p},\\
\alpha_l(G)&=\frac{A_{\alpha,l,p}}{s_J(G)}\left(\frac{D(G)}{D_0}\right)^{b_{\alpha,l,p}}.
\end{aligned}}
\tag{1}\label{eq:angle-assignment}
\end{equation}
Here \(A_{\gamma,l,p}\) and \(A_{\alpha,l,p}\) are signed amplitudes, and \(b_{\gamma,l,p}\) and \(b_{\alpha,l,p}\) are empirical degree exponents fitted to offline-optimized training angles. The mixer angles \(B_{l,p}\) are retained from the offline schedule and are not rescaled. The graph statistics are defined as
\begin{equation}
\begin{aligned}
D(G)&=\frac{2|E|}{n},\\
s_J(G)&=\sqrt{\frac{1}{|E|}\sum_{(i,j)\in E}J_{ij}^{2}}.
\end{aligned}
\tag{2}\label{eq:graph-scales}
\end{equation}
The reference degree \(D_0=38/11\approx3.4545\) is the median mean degree of the training graphs. At \(D=D_0\), the weight-normalized formula angles \(\gamma_l s_J\) and \(\alpha_l s_J\) equal their fitted amplitudes.

We use the same power-law prescription for all nontrivial instances with \(D>0\) and \(s_J>0\), with no change of rule at \(D=1\). The amplitudes and degree exponents are fitted using training graphs with \(D>1\); application at lower mean degrees extrapolates this fitted rule.

The prescription retains RMS edge-weight normalization and unscaled mixer angles from the weighted-QAOA parameter-setting approach of Sureshbabu \emph{et al.} \cite{sureshbabu2024}. For DC-QAOA, we fit separate degree powers for the cost and counterdiabatic angles at each depth and layer. Uniform positive rescaling of all edge weights gives exact inverse-weight scaling of the assigned angles, consistent with the circuit covariance in \eqref{eq:weight-covariance}. The generator second moments in \eqref{eq:generator-moments} motivate an inverse-square-root reference scale, while the exponents themselves are fitted to the training angles.

For example, the depth-1 coefficients, rounded for presentation, are
\[
\begin{aligned}
A_{\gamma,1,1}&=-0.280523,&b_{\gamma,1,1}&=-0.446584,\\
A_{\alpha,1,1}&=-0.155581,&b_{\alpha,1,1}&=-0.780388,\\
B_{1,1}&=0.479780.
\end{aligned}
\]
Substituting a target graph's \(D\) and \(s_J\) into \eqref{eq:angle-assignment} gives all three circuit angles without evaluating its quantum objective. Each depth uses its complete depth-specific coefficient set; truncating the depth-6 schedule does not provide the schedules for other depths. The coefficients are shared across graph sizes. Here, ``formula-based'' denotes direct assignment from frozen coefficients and classical graph statistics; the refinement experiment uses these assigned angles as initial conditions for a separate local search. Fig.~\ref{fig:1}(b) summarizes this procedure.

\begin{figure*}[t]
\centering
\includegraphics[width=0.95\linewidth]{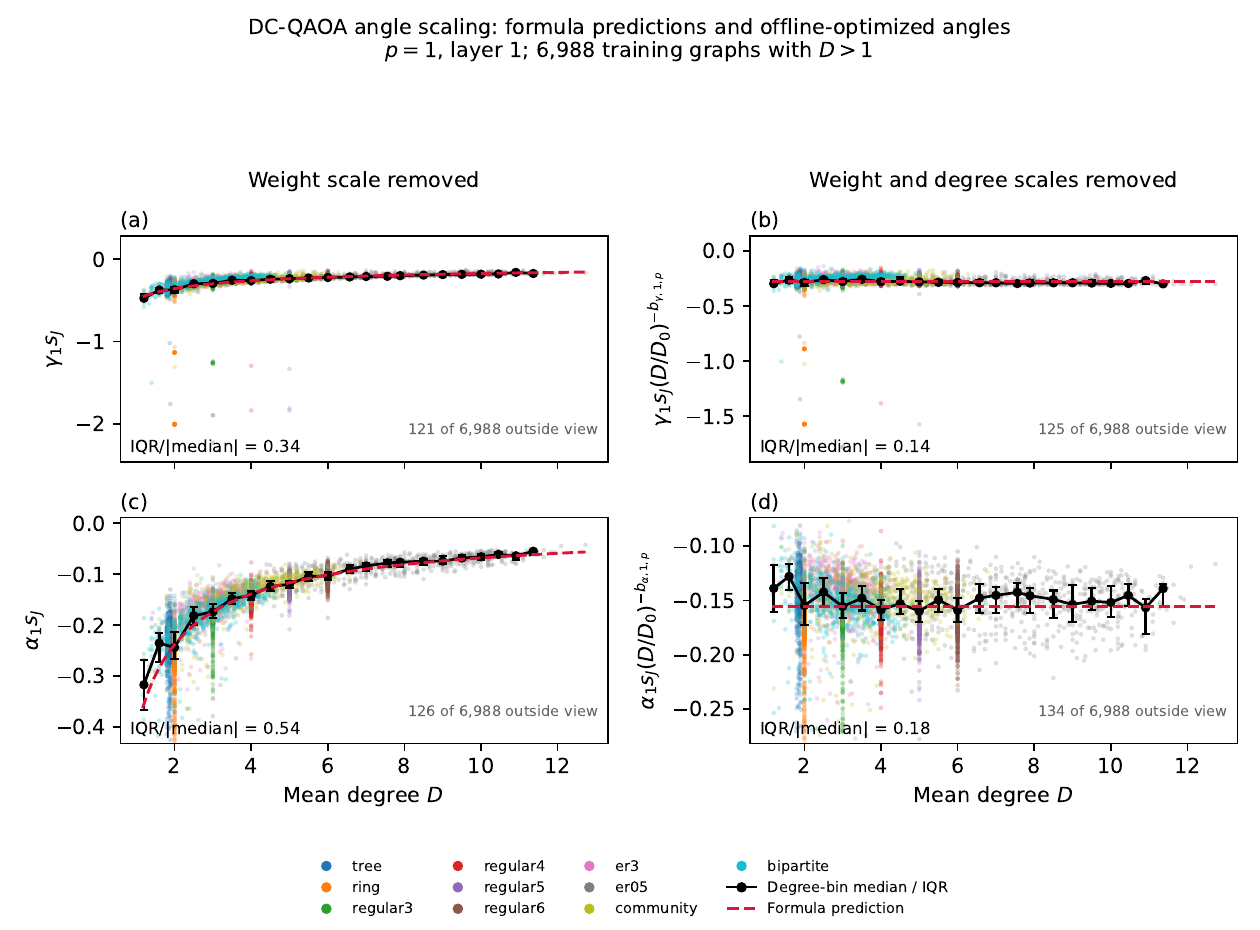}
\caption{
DC-QAOA angle scaling: formula predictions and offline-optimized angles at $p=1$ on 6988 training graphs with $D>1$. The upper and lower rows show cost and counterdiabatic angles. The left column removes the RMS-weight scale; the right column also removes the fitted degree power using \eqref{eq:rescaled-angles}. Points are colored by graph family. Black markers and error bars show degree-bin medians and interquartile ranges (IQRs) (bin width 0.5; at least 10 records). Red dashed lines show the fitted power laws on the left and their constant amplitudes on the right. IQR/$|\mathrm{median}|$ and outside-view counts use all 6988 records in the displayed sample. The 12 records with $D\leq1$ lie outside the fitting range and are excluded.}\label{fig:angle-rescaling}
\end{figure*}

We compare the power-law predictions with the first-layer offline-optimized angles at $p=1$, using the 6988 training graphs with $D>1$ (Fig.~\ref{fig:angle-rescaling}). The angles follow the sign convention in Section~\ref{offline-table-construction-and-evaluation-protocol}. The other 12 training graphs have $D\leq1$ and lie outside the fitting range.

The left-hand panels show $\gamma_1s_J$ and $\alpha_1s_J$. To remove the fitted degree dependence, we define
\begin{equation}
\begin{aligned}
\widehat\gamma_{l,p}(G)&=\gamma_l(G)s_J(G)\left(\frac{D(G)}{D_0}\right)^{-b_{\gamma,l,p}},\\
\widehat\alpha_{l,p}(G)&=\alpha_l(G)s_J(G)\left(\frac{D(G)}{D_0}\right)^{-b_{\alpha,l,p}}.
\end{aligned}
\tag{3}\label{eq:rescaled-angles}
\end{equation}
Here $b_{\gamma,l,p}$ and $b_{\alpha,l,p}$ are the fitted degree exponents. For formula-assigned angles, the rescaled variables equal the corresponding fitted amplitudes, $A_{\gamma,l,p}$ and $A_{\alpha,l,p}$. These constants are shown as horizontal red dashed lines in the right-hand panels. Applying the same transformation to the offline-optimized angles shows their dispersion about the predicted scale.

In these rescaled coordinates, the cost and counterdiabatic angles cluster around the fitted amplitudes, with relative interquartile ranges decreasing from 0.34 to 0.14 and from 0.54 to 0.18, respectively. These statistics use all 6988 records in the displayed sample, including those outside the plotted axis ranges. The rescaled pooled medians are $-0.2796$ and $-0.1552$, compared with fitted amplitudes of $-0.2805$ and $-0.1556$. Residual variation remains across individual graphs. The corresponding first-layer diagnostic at $p=4$ appears in Supplementary Fig.~S1; Supplementary Fig.~S2 reports the depth-dependent first-layer fits.

\begin{table*}[t]
\caption{
Frozen power-law coefficients in \eqref{eq:angle-assignment} for layer $l$ of a depth-$p$ circuit}\label{tab:angles}
\centering\scriptsize
\begin{tabular}{@{}ccrrrrr@{}}\toprule
$p$ & $l$ & $A_{\gamma,l,p}$ & $b_{\gamma,l,p}$ & $B_{l,p}$ & $A_{\alpha,l,p}$ & $b_{\alpha,l,p}$ \\
\midrule
1 & 1 & $-0.280523$ & $-0.446584$ & $0.479780$ & $-0.155581$ & $-0.780388$ \\
2 & 1 & $-0.229779$ & $-0.481497$ & $0.613549$ & $-0.126820$ & $-0.686905$ \\
2 & 2 & $-0.491237$ & $-0.460754$ & $0.206408$ & $-0.031658$ & $-0.247619$ \\
3 & 1 & $-0.196973$ & $-0.487714$ & $0.719497$ & $-0.114223$ & $-0.649051$ \\
3 & 2 & $-0.485915$ & $-0.583634$ & $0.331893$ & $-0.071888$ & $-0.963786$ \\
3 & 3 & $-0.383437$ & $-0.385880$ & $0.162807$ & $-0.022169$ & $-0.839589$ \\
4 & 1 & $-0.173544$ & $-0.473824$ & $0.754922$ & $-0.101909$ & $-0.618162$ \\
4 & 2 & $-0.450742$ & $-0.620655$ & $0.428698$ & $-0.069708$ & $-0.972415$ \\
4 & 3 & $-0.419815$ & $-0.589862$ & $0.253991$ & $-0.059724$ & $-1.110570$ \\
4 & 4 & $-0.364520$ & $-0.463538$ & $0.140993$ & $-0.015938$ & $-1.035730$ \\
6 & 1 & $-0.147393$ & $-0.469210$ & $0.788766$ & $-0.086443$ & $-0.587882$ \\
6 & 2 & $-0.399691$ & $-0.596859$ & $0.489933$ & $-0.062293$ & $-0.981257$ \\
6 & 3 & $-0.382752$ & $-0.644138$ & $0.348545$ & $-0.064003$ & $-1.171540$ \\
6 & 4 & $-0.371467$ & $-0.623743$ & $0.262893$ & $-0.059591$ & $-1.360612$ \\
6 & 5 & $-0.392498$ & $-0.576310$ & $0.179982$ & $-0.046145$ & $-1.422239$ \\
6 & 6 & $-0.369021$ & $-0.455031$ & $0.108804$ & $-0.008988$ & $-1.459869$ \\
\bottomrule
\end{tabular}
\end{table*}

\subsection{ Circuit definitions and scaling relations}\label{weighted-maxcut-circuit-and-scaling-analysis}

Let \(G=(V,E)\) be an \(n\)-vertex graph with strictly positive edge weights \(J_{ij}\). We minimize the Ising Hamiltonian

\begin{equation}
H_C=\sum_{(i,j)\in E}J_{ij}Z_iZ_j,
\qquad H_B=\sum_{i\in V}X_i.
\tag{4}\label{eq:ising-generators}
\end{equation}

 In \eqref{eq:ising-generators}, \(X_i\), \(Y_i\), and \(Z_i\) are Pauli operators acting on qubit \(i\). Minimizing \(H_C\) is equivalent to maximizing the weighted cut value \(C(z)=\frac12\sum_{(i,j)\in E}J_{ij}(1-z_i z_j)\) for spin assignments \(z_i\in\{-1,+1\}\). The initial state is \(|+\rangle^{\otimes n}\). A standard QAOA layer applies the cost evolution \(U_C(\gamma_l)=e^{-i\gamma_l H_C}\) followed by the mixer \(U_B(\beta_l)=e^{-i\beta_l H_B}\).

We insert an edge-weighted, first-order counterdiabatic (CD) block between these operations. The generator that motivates this block is

\begin{equation}
H_{\rm CD}=\frac{i}{2}[H_B,H_C]
=\sum_{(i,j)\in E}J_{ij}(Y_iZ_j+Z_iY_j).
\tag{5}\label{eq:cd-generator}
\end{equation}

The implemented \emph{digitized} CD block is an ordered gate product rather than a single exact exponential of the generator in \eqref{eq:cd-generator}. For each edge \((i,j)\) in lexicographic order, we apply \(e^{-i\alpha_lJ_{ij}Y_iZ_j}\) followed by \(e^{-i\alpha_lJ_{ij}Z_iY_j}\). Each layer follows the time order \(U_C(\gamma_l)\rightarrow U_{\rm CD}(\alpha_l)\rightarrow U_B(\beta_l)\), repeated for \(l=1,\ldots,p\). Fig.~\ref{fig:1}(a) shows this sequence, with the CD block representing the ordered gate product.

The gate convention is \(R_{PP}(\theta)=e^{-i\theta P\otimes P/2}\): each \(ZZ\), \(YZ\), and \(ZY\) rotation receives angle \(2J_{ij}\gamma_l\), \(2J_{ij}\alpha_l\), and \(2J_{ij}\alpha_l\), respectively, while each single-qubit \(X\) rotation receives \(2\beta_l\). This convention specifies the sign and normalization used in the numerical angle tables. Setting every \(\alpha_l=0\) recovers our QAOA circuit. At fixed \(p\), the CD extension also adds two two-qubit rotations per edge and layer; our comparison below is by \emph{layer count}, not by matched gate count.

 The weight-scale factors in \eqref{eq:angle-assignment} satisfy an exact covariance of this ordered circuit. For a fixed graph and any common positive rescaling \(J\mapsto cJ\),

\begin{equation}
\begin{aligned}
U_{cJ}\!\left(\boldsymbol\gamma/c,\boldsymbol\beta,\boldsymbol\alpha/c\right)
&=U_J(\boldsymbol\gamma,\boldsymbol\beta,\boldsymbol\alpha),\\
E_{cJ}\!\left(\boldsymbol\gamma/c,\boldsymbol\beta,\boldsymbol\alpha/c\right)
&=c\,E_J(\boldsymbol\gamma,\boldsymbol\beta,\boldsymbol\alpha),
\end{aligned}
\tag{6}\label{eq:weight-covariance}
\end{equation}
where \(E_J=\langle H_C(J)\rangle\). Each \(ZZ\), \(YZ\), and \(ZY\) rotation preserves its angle--weight product, and the mixer is independent of the edge weights. Every gate therefore remains unchanged, without requiring the CD gates to commute. Since \(s_J(cJ)=c\,s_J(J)\) and \(D\) is unchanged, the \(1/s_J\) factors in the prescription satisfy this covariance.  The normalized Ising energy, MaxCut approximation ratio, and optimal-solution probability are invariant: the final state is identical and positive weight scaling preserves the set of optimal cuts. Appendix A gives the gate-level proof of \eqref{eq:weight-covariance} and the probability relation in \eqref{eq:probability-weight-invariance}.

Pauli-string orthogonality also gives exact second moments for the traceless generators defined above:

\begin{equation}
\frac{\operatorname{Tr}(H_C^2)}{n2^n}=\frac{D}{2}s_J^2,\qquad
\frac{\operatorname{Tr}(H_{\rm CD}^2)}{n2^n}=D s_J^2.
\tag{7}\label{eq:generator-moments}
\end{equation}

The square roots of the per-vertex second moments in \eqref{eq:generator-moments} are $s_J\sqrt{D/2}$ and $s_J\sqrt D$. They define a spectral reference scale. At fixed depth and layer, \eqref{eq:angle-assignment} can be written exactly as
\begin{equation}
\boxed{\begin{aligned}
\gamma_l(G)&=\frac{C_{\gamma,l,p}}{s_J(G)\sqrt{D(G)}}h_{\gamma,l,p}(D(G)),\\
\alpha_l(G)&=\frac{C_{\alpha,l,p}}{s_J(G)\sqrt{D(G)}}h_{\alpha,l,p}(D(G)),\\
\beta_l(G)&=B_{l,p},
\end{aligned}}
\tag{8}\label{eq:common-scale}
\end{equation}
where, for $q\in\{\gamma,\alpha\}$,
\begin{equation}
\begin{aligned}
C_{q,l,p}&=A_{q,l,p}\sqrt{D_0},\\
h_{q,l,p}(D)&=\left(\frac{D}{D_0}\right)^{b_{q,l,p}+1/2}.
\end{aligned}
\tag{9}\label{eq:scale-coefficients}
\end{equation}
The residual factors retain the fitted degree dependence beyond the inverse-square-root reference scale. Equations~\eqref{eq:common-scale} and \eqref{eq:scale-coefficients} are exact for all $D>0$ and $s_J>0$, including the low-degree extrapolation range. A residual factor is constant only when its fitted exponent equals $-1/2$.

Writing either angle as the function $\theta_{q,l,p}(D,s_J)$ defined by \eqref{eq:angle-assignment}, its scaling relations are
\begin{equation}
\begin{aligned}
\theta_{q,l,p}(cD,s_J)&=c^{b_{q,l,p}}\theta_{q,l,p}(D,s_J),\\
\theta_{q,l,p}(D,cs_J)&=c^{-1}\theta_{q,l,p}(D,s_J),\quad c>0.
\end{aligned}
\tag{10}\label{eq:power-law-homogeneity}
\end{equation}
The first identity expresses homogeneity of the fitted parameter-setting function in the scalar degree argument, with its coefficients held fixed. It is not a graph transformation or an exact relation between optimal angles on different graphs. The second identity is consistent with the exact circuit covariance under uniform edge-weight rescaling in \eqref{eq:weight-covariance}.

Infinite-size calculations for the Sherrington--Kirkpatrick model show that ordinary QAOA angles can be chosen before seeing a particular signed-coupling instance \cite{farhi2022}, and large-girth regular-graph analyses connect ordinary MaxCut QAOA performance with the large-degree limit \cite{basso2022}. These results motivate degree scaling but do not determine the DC-QAOA exponents fitted here. Appendix B derives the spectral reference scale and its relation to the empirical power laws.

\subsection{Offline table construction and evaluation protocol}\label{offline-table-construction-and-evaluation-protocol}

 We use two graph cohorts spanning 10--16 and 18--22 qubits, each divided into training, validation, and test sets; the split sizes are given in Supplementary Section~S8. For each available training graph-depth pair, we optimized QAOA and DC-QAOA angles with L-BFGS-B from eight initializations and retained the lowest-energy vector for each method. These vectors are numerical search results, not certified global optima. The graph-generation rules and edge-weight sampling parameters are listed in Supplementary Section~S4.

We pool the available DC-QAOA training vectors from both cohorts: 7000 vectors at each depth $p\leq4$ and 5000 at $p=6$. We align equivalent sign branches before constructing the coefficients: if the mean cost angle of a vector is positive, we reverse all cost and mixer angles and leave the counterdiabatic angles unchanged. Because $H_C$, $H_B$, and $|+\rangle^{\otimes n}$ are real while $Y_iZ_j$ and $Z_iY_j$ are imaginary, complex conjugation maps $|\psi(\boldsymbol\gamma,\boldsymbol\beta,\boldsymbol\alpha)\rangle$ to $|\psi(-\boldsymbol\gamma,-\boldsymbol\beta,\boldsymbol\alpha)\rangle$, so the two branches give identical computational-basis probabilities. The shared reference degree $D_0$ is the median mean degree of all 7000 distinct training graphs.

For each depth, layer, and angle family $q\in\{\gamma,\alpha\}$, we fit the weight-normalized angles $\theta_{q,l}s_J$ using training graphs with $D>1$. This yields 6988 records at each depth $p\leq4$ and 4988 at $p=6$ before bin filtering. We group degrees into bins of width 0.5 by rounding $D/0.5$ to the nearest integer and retain bins containing at least 10 records. In each retained bin, $d_k$ is the median degree, $m_k$ is the median weight-normalized angle, and $n_k$ is the number of records. We fit
\[
\log|m_k|=\log|A_{q,l,p}|+b_{q,l,p}\log(d_k/D_0)
\]
by weighted least squares, assigning weight $n_k$ to each squared residual. Bins with zero angle median are excluded from the logarithmic fit. The amplitude takes the sign of the median of the retained bin medians; the retained bin medians have a common sign for every fitted depth, layer, and angle family. Cost and counterdiabatic coefficients are fitted separately.

The mixer values $B_{l,p}$ are retained from the original pooled offline schedule. That schedule used the medoid of the sign-aligned vectors $(\gamma_l s_J/f_D(D),\beta_l,\alpha_l s_J)_{l=1}^p$, where $f_D(D)=\arctan(1/\sqrt{D-1})$ for $D>1$ and $f_D(D)=1$ otherwise. The medoid is the observed normalized vector minimizing the sum of Euclidean distances to the other vectors. Only its mixer entries enter the present rule; the cost and counterdiabatic amplitudes and exponents come from the regression above.

Table~\ref{tab:angles} lists the resulting coefficients, rounded to six decimal places; evaluations use full precision. All graph sizes use the same depth-specific coefficients and $D_0$. The five depths contain 16 layers in total, giving 80 stored coefficients plus $D_0$. The exponents come from this regression rather than from a validation search over circuit performance.

For each cohort and depth, the fixed-angle QAOA comparator was selected from three candidates: a cohort-trained medoid, a transferred-angle table with RMS-weight and degree scaling, and the same table with RMS-weight scaling alone \cite{shaydulin2023,sureshbabu2024}. Selection minimizes the mean normalized Ising energy on the validation graphs; we retain these choices for the probability evaluation below. We retain the nine validation-selected comparators in Supplementary Table~S4, including the transferred-angle schedule for the smaller-size cohort at $p=6$, and share the DC-QAOA coefficients across cohorts.

The evaluation graphs are disjoint from the training pool but were examined during our earlier development of the rule. The power-law coefficients were fixed before the present evaluation and are not updated with its results.

 The main probability evaluation uses a fixed, balanced sample of 1050 test graphs: three graphs per size--family--weight cell, spanning seven sizes from 10 to 22 qubits, ten graph families, and five weight distributions. The sample-selection procedure is described in Supplementary Section~S8. We evaluate the same graphs at $p=1$--4, giving 4200 graph--depth pairs. At $p=6$, we evaluate the 600 graphs with 10--16 qubits from this subset. A separate diagnostic covers all 11 test-pool graphs with $D\leq1$ at $p=1,2,3,4,6$; one of these graphs belongs to both the main subset and its 600-graph extension. No low-degree graph is removed from either subset.

Both size ranges use the same depth-specific coefficients and reference degree. For each graph, we compute $D$ and $s_J$ from its classical description, without evaluating the target-instance objective for angle selection.  For the ideal-circuit benchmarks, we evaluate both frozen prescriptions with PennyLane \texttt{lightning.qubit}, using double-precision statevectors and \texttt{shots=None}. Before aggregation, we check graph IDs, fingerprints, depths, and the model hash.

Our primary performance measure is the probability of sampling an optimal cut. Let $C_{\max}(G)=\max_z C(z)$ and let $|z\rangle$ denote the computational-basis state associated with spin assignment $z$. For method $M$ at depth $p$, define
\begin{equation}
\begin{aligned}
\mathcal Z_{\mathrm{opt}}(G)&=\{z:C(z)=C_{\max}(G)\},\\
P_{\mathrm{opt}}^{M}(G,p)&=\sum_{z\in\mathcal Z_{\mathrm{opt}}(G)}
|\langle z|\psi_{G,p}^{M}\rangle|^2.
\end{aligned}
\tag{11}\label{eq:optimal-probability}
\end{equation}
The sum includes all degenerate optimal solutions. We compute this ideal single-measurement probability directly from the statevector and the classically enumerated optimum; the optimum is used for evaluation, not angle assignment. The paired difference is
\begin{equation}
\begin{aligned}
\Delta P_{\mathrm{opt}}(G,p)
&=P_{\mathrm{opt}}^{\mathrm{PL}}(G,p)\\
&\quad-P_{\mathrm{opt}}^{\mathrm{Q}}(G,p),
\end{aligned}
\tag{12}\label{eq:paired-probability-difference}
\end{equation}
where PL and Q denote power-law DC-QAOA and fixed-angle QAOA. Positive values favor the former; probability differences are displayed in percentage points as $100\Delta P_{\mathrm{opt}}$.

Size comparisons averaged over $p=1$--4 first take an equal-weight mean within each graph, then average over graphs. Where shown, 95\% confidence intervals are percentile intervals from 2000 graph-bootstrap resamples, stratified by size for pooled comparisons and preserving both methods and, for depth-averaged quantities, all four depths of each graph. These pointwise intervals describe variation across sampled graphs, not measurement-shot noise. Supplementary Sections~S8 and S9 provide implementation details and the depth-six results.

Offline training, the random-start optimizer benchmark, and local refinement continue to minimize energy. The latter two experiments retain the approximation ratio, $r=\mathbb E[C]/C_{\max}$, to describe their optimization trajectories; improving this expectation does not in general guarantee a higher $P_{\mathrm{opt}}$.

We also execute both frozen prescriptions on the IBM Quantum Heron processor \texttt{ibm\_fez} for 21 unit-weight 3-regular graphs, with three graphs at each size from 10 to 22 qubits, at $p=1$ and $2$. Each circuit uses 8000 shots, without target-instance angle optimization. Hardware probabilities are the observed fractions of all optimal bit strings and are analyzed separately from the ideal-circuit benchmarks. Circuits are compiled with Qiskit \cite{qiskit2024}; Supplementary Section~S3 gives the execution and sampling protocol.

\section{Results: Performance and limits of scaling-guided parameter setting}\label{results-performance-and-limits-of-scaling-guided-parameter-setting}

 We evaluate the frozen prescriptions using the optimal-solution probability and paired difference defined in \eqref{eq:optimal-probability} and \eqref{eq:paired-probability-difference}. The comparator is the validation-selected fixed-angle QAOA schedule for each size cohort and depth; neither method is optimized on the target test graph. The separate angle-search and refinement experiments use energy-based objectives and track approximation-ratio trajectories.

Section~\ref{comparison-with-fixed-angle-qaoa-across-7000-graphs} reports the primary ideal-circuit probability comparison, followed by hardware execution in Section~\ref{hardware-evaluation}. Sections III-C--III-E return to ideal-circuit evaluations of angle-search cost, refinement, and graph-dependent performance.

\subsection{ Optimal-solution probability}\label{comparison-with-fixed-angle-qaoa-across-7000-graphs}

\begin{figure}[!tbp]
\centering
\includegraphics[width=0.98\linewidth]{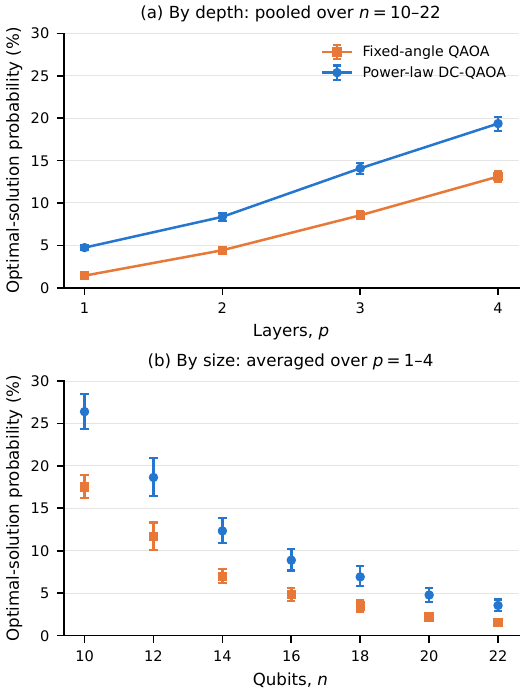}
\caption{
 Mean optimal-solution probability of power-law DC-QAOA and validation-selected fixed-angle QAOA at equal layer count. Both panels share a vertical scale. (a) Depth dependence at $p=1$--4, using the same 1050 test graphs spanning $n=10,12,14,16,18,20,22$. (b) Size dependence after equally averaging $p=1$--4 within each graph, then averaging over the 150 graphs at each size. Error bars are pointwise 95\% percentile intervals from 2000 graph-bootstrap resamples. In (a), resampling is stratified by size, preserving 150 graphs per size; in (b), graphs are resampled within the displayed size, with all four depths carried together. Both methods use the same graph resamples.}\label{fig:2}
\end{figure}

 The frozen power-law prescription yields a higher mean optimal-solution probability than fixed-angle QAOA across the four tested depths. Fig.~\ref{fig:2}(a) shows this comparison on the same 1050 test graphs at each depth, $p=1$--4. At $p=1$, the mean probabilities are 4.73\% for power-law DC-QAOA and 1.43\% for fixed-angle QAOA; at $p=4$, they reach 19.36\% and 13.11\%, respectively. The mean paired difference increases from 3.30 to 6.25 percentage points across this depth range, with pointwise 95\% paired-bootstrap intervals above zero at all four depths. Individual graphs can favor either method.

 Fig.~\ref{fig:2}(b) summarizes size dependence by averaging the four depths equally within each graph. The mean probability is higher for power-law DC-QAOA at all seven sizes. At $n=10$, the depth-averaged means are 26.39\% and 17.52\% for power-law DC-QAOA and fixed-angle QAOA; at $n=22$, they are 3.54\% and 1.55\%. For both methods, the mean probability decreases across the tested sizes.

 At $p=6$ on the 600-graph subset with 10--16 qubits, the mean optimal-solution probabilities are 36.81\% for power-law DC-QAOA and 26.69\% for fixed-angle QAOA. The mean paired difference is 10.12 percentage points, with a 95\% interval of $[9.21,11.03]$ percentage points. Supplementary Section~S9 gives the depth comparison on the same 600 graphs.

\subsection{Hardware evaluation of the frozen prescription}\label{hardware-evaluation}

\begin{figure*}[!t]
\centering
\includegraphics[width=0.98\linewidth]{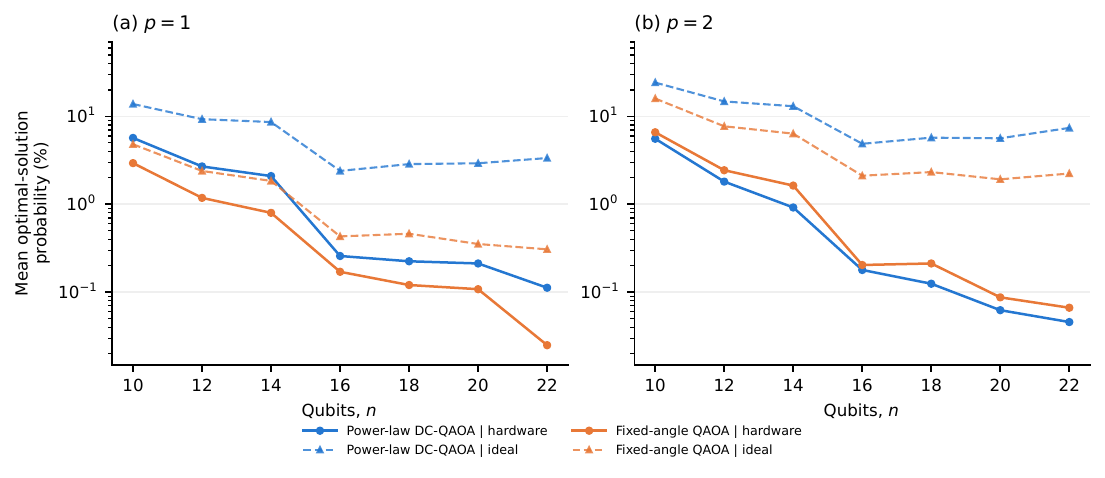}
\caption{Optimal-solution probability on \texttt{ibm\_fez} for unit-weight 3-regular graphs at (a) $p=1$ and (b) $p=2$. Each point averages three graphs at a fixed size, using 8000 shots per hardware circuit. Solid lines show raw hardware frequencies and dashed lines show ideal probabilities for the same instances and angles; blue denotes power-law DC-QAOA and orange fixed-angle QAOA. Both panels use the same logarithmic vertical scale.}\label{fig:hardware-probability}
\end{figure*}

Hardware execution tests whether the gains observed in ideal circuits persist when the frozen angles are deployed on a device. On the 21 unit-weight 3-regular graphs, mean optimal-solution probability at $p=1$ is 1.61\% for power-law DC-QAOA and 0.76\% for fixed-angle QAOA. At $p=2$, the ordering reverses, with means of 1.25\% and 1.61\%, respectively (Fig.~\ref{fig:hardware-probability}). The observed paired difference is positive on 20 of 21 graphs at $p=1$, and on four graphs at $p=2$, with one tie.

Ideal simulations of the same instances favor power-law DC-QAOA at both depths. At $p=2$, the mean transpiled two-qubit gate counts are 479.7 for power-law DC-QAOA and 232.6 for QAOA (Supplementary Fig.~S4). The larger two-qubit gate count may contribute to the reversal. Because all instances have $D=3$ and unit weights, this experiment tests hardware deployment rather than the degree and weight scaling. Supplementary Section~S3 reports the paired results and circuit-resource statistics.

\subsection{Reduction in target-instance angle-search cost}\label{reduction-in-target-instance-angle-search-cost}

 To assess target-instance angle-search cost, we compare direct formula assignment with variational energy minimization of the same DC-QAOA circuit on a separate set of 12 held-out graphs at \(n=12\) and \(p=2\). We track the approximation ratio during this energy search. The set contains three graphs for each combination of graph structure (3-regular or Erd\H{o}s--R\'enyi with mean degree 3, ER3) and edge weights (unit or uniformly sampled positive weights).

We simulate the circuits with PennyLane \cite{bergholm2018} \texttt{lightning.qubit} and run L-BFGS-B, COBYLA, and Adam from the same 20 random initial angle vectors per graph, yielding 240 runs per optimizer.  Each run allows at most 500 evaluations of the expected Ising energy, including finite-difference and line-search calls where applicable. At evaluation \(k\), each optimizer's trajectory is the mean across runs of the best expected approximation ratio observed up to \(k\). Runs that end early retain their last best value. Initialization, optimizer settings, and evaluation counting are specified in Supplementary Section~S7.

\begin{figure}[!tbp]
\centering
\includegraphics[width=0.98\linewidth]{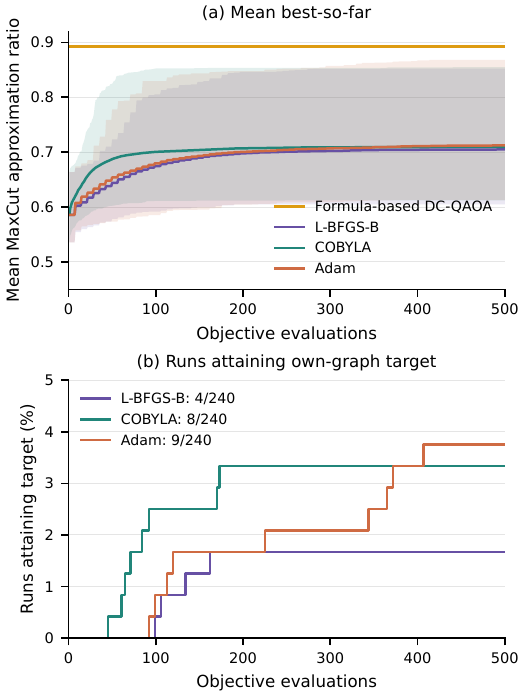}
\caption{
 Formula-based DC-QAOA versus target-instance energy minimization on 12 graphs at $n=12$ and $p=2$. (a) Mean best-so-far approximation-ratio trajectories, with 10th–90th percentile bands across individual runs and the power-law formula mean shown as a fixed reference. The horizontal line is the 12-graph mean; attainment in (b) is assessed against each graph's formula-based result.  (b) Cumulative percentage of all 240 runs per optimizer whose best-so-far approximation ratio reaches their graph-specific formula value minus 0.001; the vertical axis spans 0\%--5\%, and the legend reports attained counts. Each run has at most 500 actual objective evaluations. Runs that end early retain their last best value in (a); those that stop before reaching their graph-specific target remain below that target in (b).}\label{fig:3}
\end{figure}

The formula-based angles attain a mean approximation ratio of 0.8921 on these graphs without target-instance objective calls for angle selection. At the 500-evaluation budget, the mean best-so-far ratios are 0.7049 for L-BFGS-B, 0.7096 for COBYLA, and 0.7124 for Adam. All three mean trajectories remain below the formula-based mean throughout this budget (Fig.~\ref{fig:3}(a)).

 Within this budget, 4/240 L-BFGS-B runs, 8/240 COBYLA runs, and 9/240 Adam runs attain a best-so-far approximation ratio at least as high as their graph-specific formula value minus 0.001 (Fig.~\ref{fig:3}(b)).

\subsection{Refinement from formula-assigned angles}\label{refinement-from-formula-assigned-angles}

 To assess whether formula-assigned angles provide useful initial conditions for energy minimization, we reuse the 12 held-out graphs at \(n=12\) and \(p=2\) from Section~\ref{reduction-in-target-instance-angle-search-cost}. For each graph, we run L-BFGS-B from the formula-assigned angles and from 20 random initializations, using double-precision simulation and adjoint gradients.  Each value-and-gradient call returns the expected Ising energy and its adjoint gradient. These calls differ from the energy-only evaluations counted in Section~\ref{reduction-in-target-instance-angle-search-cost}; the random-initialization distribution also differs between the two protocols. Supplementary Section~S6 gives the detailed protocol.

\begin{figure}[!tbp]
\centering
\includegraphics[width=0.98\linewidth]{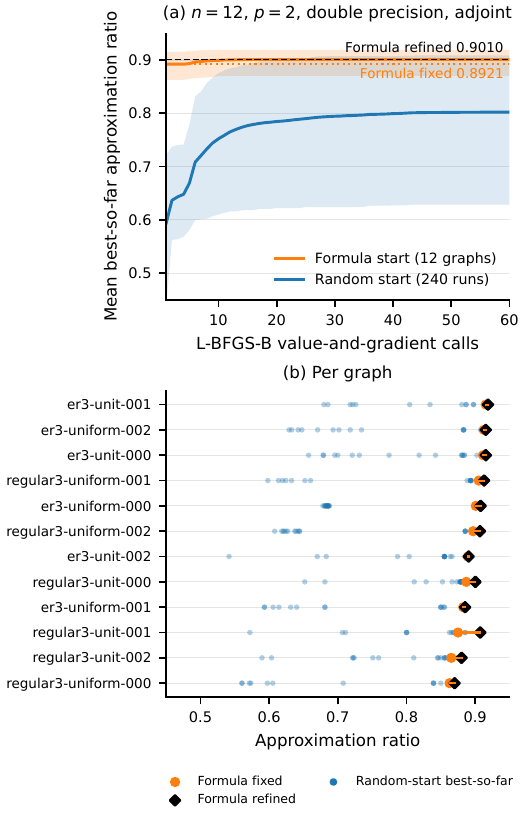}
\caption{
 Energy-based refinement from formula-assigned angles on the same 12 held-out graphs used in Fig.~\ref{fig:3}, with $n=12$ and $p=2$. (a) Mean best-so-far approximation ratio versus L-BFGS-B value-and-gradient calls for formula initialization (orange; 12 runs) and random initialization (blue; 20 runs per graph, 240 in total). The orange band shows the full range over graphs; the blue band shows the 10th--90th percentiles over random-start runs. Runs that terminate early retain their last best value.  Panel (a) shows the first 60 value-and-gradient calls; panel (b) uses the best-so-far values at termination of the complete runs. The dotted line marks the mean ratio at the formula-assigned angles; the dashed line marks the mean best-so-far ratio at termination of the formula-initialized runs.  (b) Graph-specific ratios for direct formula assignment (orange circles), and best-so-far ratios at termination for formula-initialized refinement (black diamonds) and all 20 random-start runs per graph (blue points). Optimizer settings are given in Supplementary Section~S6.}\label{fig:formula-refinement}
\end{figure}

 Formula-initialized refinement raises the mean best-so-far approximation ratio from 0.8921 to 0.9010, an absolute gain of 0.0089, using an average of about 20 value-and-gradient calls per run (Fig.~\ref{fig:formula-refinement}). The graph-specific gains range from 0.0011 to 0.0323, with a median of 0.0072.  For each graph, the best-so-far trajectory reaches 99\% of its gain from initialization to termination within 8--17 calls.

Under this refinement protocol, the 240 random-start runs attain a mean best-so-far ratio of 0.8027 at termination. Of these runs, 70 reach or exceed their graph's unrefined formula value within a tolerance of 0.001. On 10 of the 12 graphs, the formula-initialized result agrees within $10^{-4}$ with the best value found among the 20 random-start runs; on the remaining two, the best random-start value is higher by approximately 0.001 or less.  These results support using formula-assigned angles to initialize local energy minimization and improve the expected cut value on the tested instances.

\subsection{Dependence on graph structure and edge weights}\label{dependence-on-graph-structure-and-edge-weights}

\begin{figure*}[!t]
\centering
\includegraphics[width=0.98\textwidth]{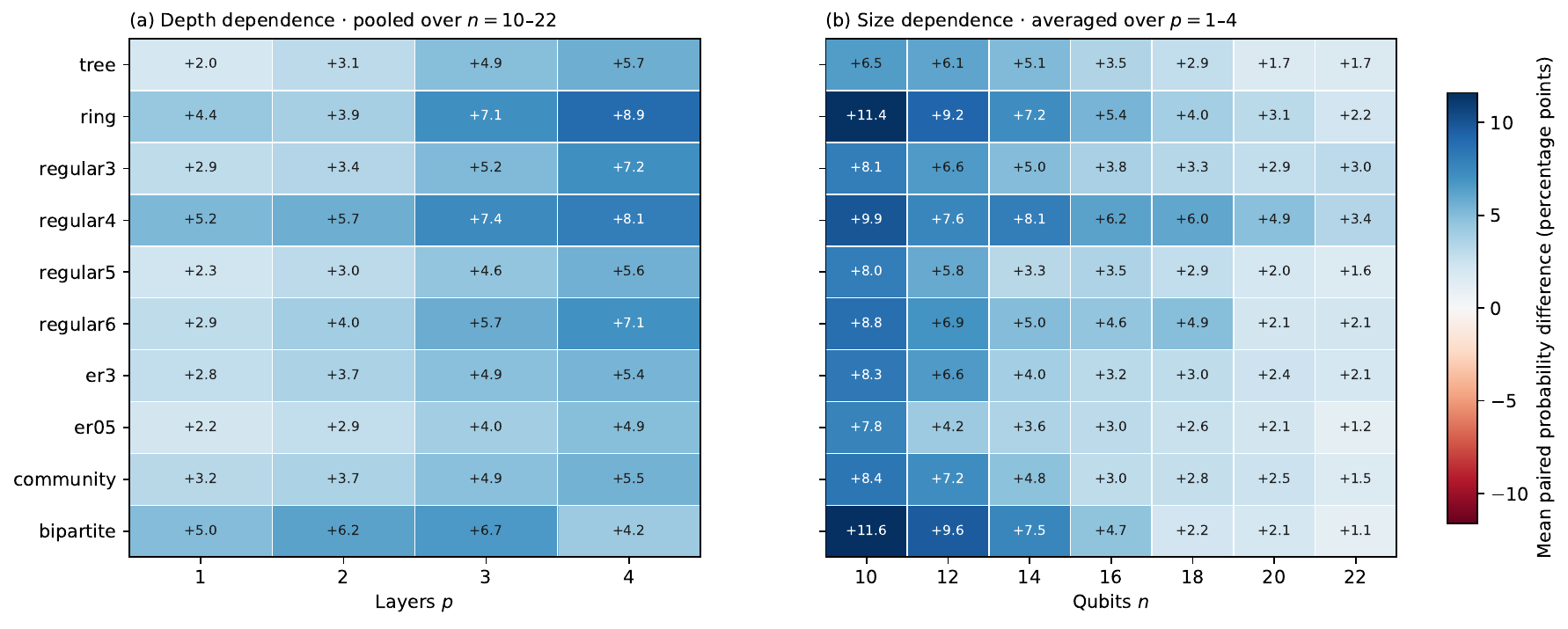}
\caption{
 Mean paired optimal-solution probability difference, $100\Delta P_{\mathrm{opt}}$, in percentage points, with $\Delta P_{\mathrm{opt}}$ defined in \eqref{eq:paired-probability-difference}. Positive values favor power-law DC-QAOA. (a) Ten graph families across $p=1$--4, pooled over $n=10$--22; each cell contains 105 paired graphs. (b) The same families across $n=10,12,14,16,18,20,22$; each cell first averages the paired differences equally over $p=1$--4 within each graph, then averages over 15 graphs. Both panels share a zero-centered color scale, with blue indicating positive differences and red indicating negative differences. Annotations show means rounded to one decimal place. Graph-generation rules are given in Supplementary Section~S4.}\label{fig:4}
\end{figure*}

 The mean probability advantage varies across graph families and remains positive in both aggregated views of Fig.~\ref{fig:4}. Panel (a) shows the results for each family by depth, pooling over $n=10$--22 with 105 paired graphs per cell. All 40 family--depth means are positive. Panel (b) shows the results for each family by size after averaging $p=1$--4 equally within each graph, then averaging over the 15 graphs in each family--size cell. All 70 depth-averaged means are positive.

These averages can conceal depth-specific reversals. Among the 280 family--size--depth cells, the bipartite ensemble at $n=22$ and $p=4$ has a mean paired difference of $-0.048$ percentage points, whereas its four-depth average is $+1.052$ percentage points. Seven of the 15 graphs in this cell favor QAOA.

\begin{figure}[!tbp]
\centering
\includegraphics[width=0.98\linewidth]{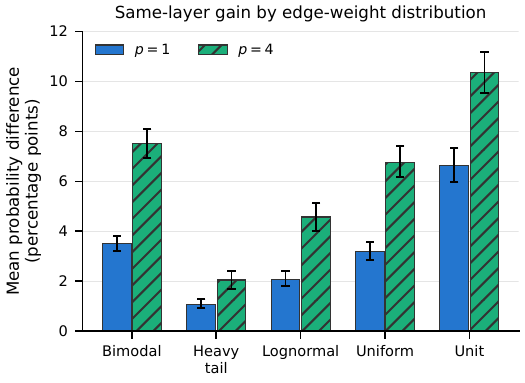}
\caption{
 Mean paired optimal-solution probability difference, $100\Delta P_{\mathrm{opt}}$, in percentage points, for five positive edge-weight distributions at $p=1$ and $p=4$. Each bar uses 210 paired graph results pooled over $n=10,12,14,16,18,20,22$ and ten graph families, with 30 graphs per size. Error bars show pointwise 95\% percentile intervals from 2000 graph-bootstrap resamples of the paired differences. Resampling is stratified by size, preserves 30 graphs per size, and keeps both methods paired on each graph. The distributions, from left to right, are bimodal, heavy-tail, lognormal, uniform, and unit weights; sampling parameters are given in Supplementary Section~S4. Blue bars show $p=1$; green hatched bars show $p=4$. Positive values favor power-law DC-QAOA.}\label{fig:5}
\end{figure}

 The magnitude of the probability advantage also depends on the edge-weight distribution. Fig.~\ref{fig:5} compares five positive-weight distributions at $p=1$ and $p=4$, with 210 paired graphs per bar pooled over $n=10$--22 and ten graph families. All five distributions have positive mean paired probability differences at both depths, with pointwise 95\% paired-bootstrap intervals above zero.

The heavy-tail group has the smallest mean advantage at both displayed depths: 1.09 percentage points at $p=1$ and 2.05 at $p=4$. Unit weights give the largest mean differences, 6.64 and 10.35 percentage points, respectively. Each weight group has a larger mean paired probability difference at $p=4$ than at $p=1$.

\section{Discussion}\label{discussion}

Across the balanced sample of 1050 graphs at $p=1$--4, the frozen power-law prescription yields a higher mean optimal-solution probability than validation-selected fixed-angle QAOA at each equal layer count. The mean paired probability difference increases from 3.30 to 6.25 percentage points across these depths. On the same 600 smaller graphs evaluated at $p=1,2,3,4,6$, the mean advantage reaches 10.12 percentage points at $p=6$ (Supplementary Section~S9). These evaluations use parameters derived from offline energy optimization; neither the power-law coefficients nor the QAOA comparator choices were selected to maximize probability on these test graphs.

The energy-based experiments address two practical questions: how much angle search the formula saves, and whether it gives a good starting point for refinement. On the separate 12-graph benchmark, the mean best-so-far approximation-ratio trajectories remain below the formula-based mean within the tested 500-energy-evaluation budget. Formula-initialized refinement raises the mean best-so-far ratio from 0.8921 to 0.9010 using about 20 value-and-gradient calls. These improvements concern expected cut value, not necessarily optimal-solution probability. On the same 1050 graphs, the mean approximation-ratio advantage decreases from 0.0659 at $p=1$ to 0.0111 at $p=4$, even as the mean paired probability difference increases (Supplementary Section~S10). The two metrics probe different properties of the output distribution.

The prescription moves angle optimization to an offline stage and assigns target-instance angles from graph statistics and frozen coefficients. Circuit evaluations are still needed to estimate an objective or sample a solution. Uniform edge-weight scaling preserves the output state and optimal-solution probability under the corresponding angle transformation (Appendix A), while the degree powers are fitted separately to cost and counterdiabatic training angles at each depth and layer. In the first-layer diagnostics at $p=1$ and $p=4$, the rescaled angles cluster around common amplitudes, with residual variation across graphs.

The probability advantage varies across graph and weight classes. All displayed family aggregates and the five weight-group means at $p=1$ and $p=4$ are positive, but averaging can conceal losses: the bipartite group at $n=22$, $p=4$ has a mean difference of $-0.048$ percentage points, although its four-depth average is positive (Supplementary Section~S10). Individual graphs can also favor QAOA. The heavy-tail group has the smallest mean advantage among the five displayed weight groups at both depths. This is consistent with the instance dependence reported for QAOA initialization strategies \cite{katial2025}. Local tree-like neighborhoods have also been used to compute and transfer QAOA parameters in random regular graphs \cite{wybo2025}, a different setting from our tree-family test.

The same formula is defined for every positive mean degree, but its coefficients are fitted using $D>1$. On the 11-graph low-degree diagnostic, the mean probability difference is positive at $p=1,2,3$ and negative at $p=4,6$, reaching $-8.74$ and $-15.62$ percentage points, respectively (Supplementary Section~S5). The rule should therefore not be applied below $D=1$ without further checks. Because the evaluation pool had been examined during the development of the rule, these results are not a fully blind test.

The ideal-circuit and hardware comparisons use equal layer counts. Each DC-QAOA layer adds two-qubit rotations, so equal layer counts do not imply matched entangling-gate budgets. Liu and Wang compared a nested-commutator DC-QAOA ansatz with QAOA at equal total CNOT counts and found that their relative performance depends on system size \cite{liuwang2025}. Their study addresses the resource efficiency of the ansatz, whereas the present work evaluates a frozen parameter-setting prescription without target-instance angle optimization. The hardware results depend on depth: the mean probability advantage persists at $p=1$ but reverses at $p=2$ on the tested unit-weight 3-regular graphs. Matched-gate-budget comparisons and broader hardware evaluations are needed to establish the scope of this behavior.

\section{Conclusion}\label{conclusion}

We develop empirical degree-scaling laws for DC-QAOA parameter setting, combined with exact covariance under uniform positive edge-weight rescaling. After the fitted weight and degree scales are removed, the offline-optimized angles cluster around common amplitudes, and the resulting frozen prescription assigns parameters without target-instance variational optimization. On 1050 graphs at $p=1$--4 and the 600-graph $p=6$ extension, the prescription improves mean optimal-solution probability over the selected fixed-angle QAOA comparators at equal layer count. In separate energy-based tests, the formula angles give a higher mean approximation ratio than random-start optimization reaches within 500 energy evaluations and provide good starting points for local refinement.

The probability advantage holds in aggregate, while individual instances can show losses and the low-degree diagnostic has negative mean differences at higher depths. These findings support scaling-based parameter assignment within the tested graph and depth ranges; hardware execution on unit-weight 3-regular graphs retains a mean advantage at $p=1$ but reverses the ordering at $p=2$. Performance at matched gate budgets and beyond the tested graph and device settings remains to be evaluated.

\section*{Code and data availability}\label{code-and-data-availability}

The code is publicly available on GitHub at \url{https://github.com/yaochengchen/Scaling_DCQAOA}. The original data are available on Zenodo at doi: \href{https://doi.org/10.5281/zenodo.23630629}{10.5281/zenodo.23630629}.

\appendices
\begin{samepage}
\section{Circuit covariance under uniform edge-weight scaling}\label{appendix-a.-circuit-covariance-under-uniform-edge-weight-scaling}

\subsection{Proposition and assumptions}\label{a.1.-proposition-and-assumptions}

Fix a graph \(G=(V,E)\) with a nonempty edge set, positive weights \(J=(J_{ij})_{(i,j)\in E}\), circuit depth \(p\), an initial state, and the gate and edge order specified in Section~\ref{weighted-maxcut-circuit-and-scaling-analysis}. For any \(c>0\), set \((cJ)_{ij}=cJ_{ij}\) and transform all layer parameters to \((\boldsymbol\gamma/c,\boldsymbol\beta,\boldsymbol\alpha/c)\). Let \(U_J\) denote the unitary of the implemented edgewise gate sequence. Then
\end{samepage}

\[
U_{cJ}\!\left(\boldsymbol\gamma/c,\boldsymbol\beta,
\boldsymbol\alpha/c\right)
=U_J(\boldsymbol\gamma,\boldsymbol\beta,\boldsymbol\alpha).
\tag{A1}
\]

Equation (A1) is an operator identity for the specified \emph{digitized circuit}. It does not identify the ordered product of CD gates with the single exponential \(e^{-i\alpha_lH_{\rm CD}}\).

\subsection{Gate-level proof}\label{a.2.-gate-level-proof}

Because \(H_C(cJ)=cH_C(J)\), the cost evolution in layer \(l\) obeys

\[
e^{-i(\gamma_l/c)H_C(cJ)}
=e^{-i\gamma_lH_C(J)}.
\tag{A2}
\]

For every edge \((i,j)\), the two CD rotations obey

\[
\begin{aligned}
e^{-i(\alpha_l/c)(cJ_{ij})Y_iZ_j}
&=e^{-i\alpha_lJ_{ij}Y_iZ_j},\\
e^{-i(\alpha_l/c)(cJ_{ij})Z_iY_j}
&=e^{-i\alpha_lJ_{ij}Z_iY_j}.
\end{aligned}
\tag{A3}
\]

The mixer \(H_B=\sum_iX_i\) contains no edge weights and \(\beta_l\) is unchanged. Every gate therefore remains unchanged under rescaling. Their ordered product is identical, proving (A1) for each layer and the full circuit without exchanging CD gates or assuming commutativity.

For completeness, \([X_i,Z_i]=-2iY_i\) gives

\[
\frac{i}{2}[H_B,H_C(J)]
=\sum_{(i,j)\in E}J_{ij}(Y_iZ_j+Z_iY_j)
=H_{\rm CD}(J).
\tag{A4}
\]

Thus \(H_{\rm CD}(cJ)=cH_{\rm CD}(J)\). This relation connects the selected rotations to their motivating generator; the gate-level proof follows directly from the invariance of each angle--weight product.

\subsection{Energy, optimal-solution probability, and scope}\label{a.3.-energy-approximation-ratio-and-scope}

For the same weight-independent initial state, (A1) implies identical final states and computational-basis measurement probabilities. Since \(H_C(cJ)=cH_C(J)\), the expected Ising energy satisfies

\[
E_{cJ}\!\left(\boldsymbol\gamma/c,\boldsymbol\beta,
\boldsymbol\alpha/c\right)
=c\,E_J(\boldsymbol\gamma,\boldsymbol\beta,\boldsymbol\alpha).
\tag{A5}
\]

The normalization factor \(\sum_eJ_e\) also scales by \(c\), leaving \(E_J/\sum_eJ_e\) unchanged. Every weighted MaxCut value obeys \(C_{cJ}(z)=cC_J(z)\), so the exact maximum cut value scales by \(c\) as well. Because the measurement probabilities are identical, the expected cut value also scales by \(c\); therefore the approximation ratio is unchanged.

The optimal-solution probability in \eqref{eq:optimal-probability} is invariant as well. Let $\mathcal Z_{\mathrm{opt}}(J)=\operatorname*{arg\,max}_{z}C_J(z)$. Since $C_{cJ}(z)=cC_J(z)$ and $c>0$, we have $\mathcal Z_{\mathrm{opt}}(cJ)=\mathcal Z_{\mathrm{opt}}(J)$, including every degenerate optimum. Equation (A1) preserves the final state for the same initial state, so
\[
\begin{aligned}
P_{\mathrm{opt},cJ}\!\left(\boldsymbol\gamma/c,\boldsymbol\beta,
\boldsymbol\alpha/c\right)
&=P_{\mathrm{opt},J}\!\left(\boldsymbol\gamma,\boldsymbol\beta,
\boldsymbol\alpha\right).
\end{aligned}
\tag{A6}\label{eq:probability-weight-invariance}
\]
Here $P_{\mathrm{opt},J}$ sums the computational-basis probabilities over $\mathcal Z_{\mathrm{opt}}(J)$. Thus, both the event being measured and its probability are unchanged.

Under the same rescaling, \(D(cJ)=D(J)\) and \(s_J(cJ)=c\,s_J(J)\). The prescription in Section~\ref{formula-based-angle-setting-with-graph-dependent-rescaling} consequently gives \(\boldsymbol\gamma(cJ)=\boldsymbol\gamma(J)/c\), \(\boldsymbol\beta(cJ)=\boldsymbol\beta(J)\), and \(\boldsymbol\alpha(cJ)=\boldsymbol\alpha(J)/c\), as required by (A1).

This covariance concerns a common rescaling of all edge weights. The weight statistic, degree factors, table angles, and transfer under changes in relative edge weights require separate justification or empirical evaluation. A correspondence between sets of globally optimal angles additionally requires the feasible parameter domains to map onto one another under the angle transformation. Equation (A1) itself imposes no such domain condition.

\section{Generator second moments and the common degree scale}\label{appendix-b.-generator-second-moments-and-the-common-degree-scale}

\subsection{Exact second-moment identities}\label{b.1.-exact-second-moment-identities}

Assume a simple undirected graph with a nonempty edge set, each edge counted once. For the traceless \(H_C\) and \(H_{\rm CD}\) in Section~\ref{weighted-maxcut-circuit-and-scaling-analysis}, distinct \(n\)-qubit Pauli strings are orthogonal under the normalized Hilbert--Schmidt inner product, \(2^{-n}\operatorname{Tr}(P_aP_b)=\delta_{ab}\). The strings \(Z_iZ_j\) are distinct across edges. The strings \(Y_iZ_j\) and \(Z_iY_j\) are distinct both within each edge and across different edges. Each Pauli string squares to the identity, and the cross terms have zero trace. Therefore,

\[
\frac{\operatorname{Tr}(H_C^2)}{2^n}=\sum_{(i,j)\in E}J_{ij}^2,
\qquad
\frac{\operatorname{Tr}(H_{\rm CD}^2)}{2^n}=2\sum_{(i,j)\in E}J_{ij}^2.
\tag{B1}
\]

With \(D=2|E|/n\) and \(s_J^2=|E|^{-1}\sum_{(i,j)\in E}J_{ij}^2\), (B1) becomes

\[
\frac{\operatorname{Tr}(H_C^2)}{n2^n}=\frac{D}{2}s_J^2,
\qquad
\frac{\operatorname{Tr}(H_{\rm CD}^2)}{n2^n}=D s_J^2.
\tag{B2}
\]

These are exact infinite-temperature second moments of the specified traceless generators. For the MaxCut objective in Section~\ref{weighted-maxcut-circuit-and-scaling-analysis}, removing the identity term from its operator form leaves \(-H_C/2\), whose second moment is one quarter of that of the Ising generator in (B1). Both the identity term and the multiplicative normalization must therefore be accounted for when changing conventions. For parallel edges, the coefficients of identical Pauli strings must be combined before the second moments are evaluated.

\subsection{Relation to the empirical angle rule}\label{b.2.-relation-to-the-empirical-angle-rule}

The square roots of the per-vertex second moments in (B2) define the spectral-scale variables $\widetilde\gamma_l=\gamma_l s_J\sqrt{D/2}$ and $\widetilde\alpha_l=\alpha_l s_J\sqrt D$. These differ from the fitted-power normalization in \eqref{eq:rescaled-angles}.

For $q\in\{\gamma,\alpha\}$, direct substitution of the power-law rule gives
\[
\begin{aligned}
\theta_{q,l,p}(G)
&=\frac{A_{q,l,p}}{s_J}\left(\frac{D}{D_0}\right)^{b_{q,l,p}}\\
&=\frac{A_{q,l,p}\sqrt{D_0}}{s_J\sqrt D}
\left(\frac{D}{D_0}\right)^{b_{q,l,p}+1/2}.
\end{aligned}
\tag{B3}
\]
The equality follows from $\sqrt{D_0/D}(D/D_0)^{1/2}=1$ and holds for every $D>0$ and $s_J>0$; it is not a large-degree expansion. Multiplication by the corresponding spectral strength yields
\[
\begin{aligned}
\widetilde\gamma_l&=A_{\gamma,l,p}\sqrt{D_0/2}
\left(\frac{D}{D_0}\right)^{b_{\gamma,l,p}+1/2},\\
\widetilde\alpha_l&=A_{\alpha,l,p}\sqrt{D_0}
\left(\frac{D}{D_0}\right)^{b_{\alpha,l,p}+1/2}.
\end{aligned}
\tag{B4}
\]
At fixed depth and layer, these variables are independent of $D$ only if the corresponding exponent is $-1/2$ (for nonzero amplitude). For a fixed fitted rule extended to $D\to\infty$, their magnitudes tend to zero when $b<-1/2$ and grow when $b>-1/2$; they remain constant when $b=-1/2$. These are algebraic properties of the extended rule, not predictions of its accuracy outside the training range. In contrast, the fitted-power normalization in \eqref{eq:rescaled-angles} removes the full prescribed degree dependence and returns the amplitudes $A_{\gamma,l,p}$ and $A_{\alpha,l,p}$ exactly for formula-assigned angles.

Equations (B1) and (B2) establish the spectral reference scale; (B3) and (B4) describe how the empirical rule relates to that scale. The second moments do not encode the initial state, graph structure beyond $D$ and $s_J$, noncommuting gate effects, or the objective landscape. In particular, the implemented ordered CD gate product is generally not $e^{-i\alpha_lH_{\rm CD}}$. The fitted degree exponents therefore require evidence from the training angles and evaluation on other graphs, rather than following from the moment identities alone.

\section*{Acknowledgment}\label{acknowledgment}

Generative artificial intelligence (AI) tools (OpenAI ChatGPT and Anthropic Claude) were used to assist with software development and language editing; the authors reviewed and verified all AI-assisted output and are responsible for the manuscript and the software. Y. Chen acknowledges the use of IBM Quantum services for this work. The views expressed are those of the author, and do not reflect the official policy or position of IBM or the IBM Quantum team.

Author contributions are as follows.

Y. Chen: Conceptualization; methodology; software; data curation; writing---review and editing.

J. Zhuang: Conceptualization; methodology; validation; investigation; writing---original draft; writing---review and editing.